\documentclass[12pt]{article} 

\usepackage{amssymb,amsmath,epsfig,color}
\usepackage{framed}
\usepackage{epstopdf}
\newcommand{\rem}[1]{}
\newcommand{\remfigure}[1]{#1}

\newcommand{\pa}{\partial}

\newcommand{\bS}{\boldsymbol{S}}

\newcommand{\bq}{\boldsymbol{q}}

\newcommand{\bx}{\boldsymbol{x}}

\newcommand{\bom}{\boldsymbol{\omega}}

\newcommand{\bhz}{\mathbf{\hat{z}}}

\rem{
\DeclareMathAlphabet{\mathbi}{OML}{cmm}{b}{it}

}
 
\def\contract{\makebox[1.2em][c]{\mbox{\rule{.6em}
{.01truein}\rule{.01truein}{.6em}}}}
\newcommand{\bfi}{\bfseries\itshape}
\newcommand{\bfib}{\color{blue}\bfseries\itshape}
\newcommand{\comment}[1]{\vspace{1 mm}\par 
\marginpar{\large\underline{}}\noindent
\framebox{\begin{minipage}[c]{0.90 \textwidth}
{\bfib #1} \end{minipage}}\vspace{1 mm}\par}
\newtheorem{theorem}{Theorem}
\newtheorem{proposition}[theorem]{Proposition}
\newtheorem{remark}[theorem]{Remark}
\newenvironment{proof}[1][Proof]{\textbf{#1.} }{\ \rule{0.5em}{0.5em}}

\begin{document}
 
\title{
Geometric gradient-flow dynamics with singular solutions}
\author{\vspace{2mm} \small
Darryl D. Holm$^{\,\text{\scriptsize 1,\,2}}$, Vakhtang Putkaradze$^{\,\text{\scriptsize
3,\,4}}$ and Cesare Tronci$^{\,\text{\scriptsize 1,\,5}}$ \\
\hspace{-0.9cm}
\footnotesize $^1$ \sl Department of Mathematics, Imperial College London, London
SW7 2AZ, UK\\
\hspace{-0.9cm}
\footnotesize $^2$ \sl Computer and Computational Science Division,
\footnotesize\it Los Alamos National Laboratory, Los Alamos, NM, 87545 USA
\\
\hspace{-0.9cm}
\footnotesize $^3$ \sl Department of Mathematics, Colorado State University,
Fort Collins, CO 80523 USA\\
\hspace{-0.9cm}
\footnotesize $^4$ \sl 
Department of Mechanical Engineering, University of New Mexico, 
Albuquerque NM 87131\\
\hspace{-0.9cm}
\footnotesize $^5$ \sl TERA Foundation for Oncological Hadrontherapy, 
11 V. Puccini, Novara 28100, Italy\\ \\
}

\normalsize

\date{}
\maketitle
\vspace{-1.15cm}
\begin{abstract} \noindent
The gradient-flow dynamics of an arbitrary geometric quantity is derived using a generalization of Darcy's Law. We consider flows in both Lagrangian and Eulerian formulations. The Lagrangian formulation includes a dissipative modification of fluid mechanics. Eulerian equations for self-organization of scalars, 1-forms and 2-forms  are shown to reduce to nonlocal characteristic equations. 
We identify singular solutions of these equations corresponding to  collapsed (clumped) states and discuss their evolution. 



\end{abstract}

\vspace{-0.45cm}
\tableofcontents

\section{Introduction} 

\paragraph{Singular solutions in continuum models.} Under certain conditions, advection equations in continuum dynamics support singular $\delta$-like solutions. The simplest example is the conservation equation for the mass density $\rho({\bf x},t)$ of the form 
\[
\pa_t\rho+{\rm div}(\rho\,{\bf u})=0
\,.
\]
As one may verify, this equation admits the single-particle trajectory 
\[
\rho({\bf x},t)=w\,\delta({\bf x - Q}(t))
\]
as its solution, for a constant weight $w$, provided the velocity vector field satisfies ${\bf u}({\bf Q}(t),t)={\bf \dot{Q}}(t)$ and is sufficiently smooth. 

More generally, singular solutions exist for {\it any} tensor
quantity undergoing the flow of a smooth vector field. One writes 
\[
\pa_t {\sf T}+\pounds_{\bf u}{\sf T}=0
\,,
\]
where ${\sf T}$ is a generic tensor field and $\pounds_{\bf u}$ stands for the Lie derivative with respect to the vector field ${\bf u}$, so that the corresponding term represents fluid transport \cite{MarsdenRatiu}. The singular solutions
for this equation may be written in general as
\[
{\sf T}({\bf x},t)=\sum_{i=1}^N\int \!{\sf P}_{\!i}(s,t)\,\delta({\bf x-Q}_i(s,t))\,{\rm d}s
\,,
\]
where the sum refers to $N$ different singular solutions and $s$ is a coordinate on the $i$-th submanifold in $\mathbb{R}^3$ (a curve or
surface). A famous example of this more general case is provided by Euler's vorticity equation for ideal
fluids: 
\[
\pa_{t\,}\bom=-\pounds_{\bf u}\,\bom={\rm curl}(\bom\times{\bf u})
\,,
\]
with nonlocal velocity ${\bf u}={\rm curl}^{-1}\,\bom$. In this case the tensor quantity $\bom$ is an exact two-form and the singular vortex solutions for this system are supported either on a curve (in the case of filaments) or on a surface (in the case of sheets) in $\mathbb{R}^3$ \cite{Sa1992}. In two dimensions, one has point vortices instead.

\paragraph{Emergent singularities in hydrodynamics.} 
Another interesting example of this kind of advection equation is provided by the dispersionless Camassa-Holm (CH) equation in one dimension \cite{CaHo1993} or its generalization in higher dimensions, the EPDiff equation \cite{HoMa2004}. The dispersionless CH and EPDiff equations possess singular solutions that emerge {\it spontaneously} from {\it any} smooth initial distribution. In particular, these are advection equations for the fluid momentum represented in geometric terms by a one-form density $m \otimes{\rm d}V={\bf m}\cdot{\rm d}{\bf x}\otimes{\rm d}V$. Just as in the case of vorticity dynamics, the momentum conservation equation 
\[
\Big(\pa_t+\pounds_{\bf u} \Big)\,(m \otimes{\rm d}V)=0
\]
involves a nonlocal velocity vector field ${\bf u}=G*{\bf
m}$, where $G$ is the Green's function for a positive definite symmetric operator, typically chosen to be $G=(1-\Delta)^{-1}$. The singular solutions of CH and EPDiff emerge spontaneously from any confined smooth initial distribution of velocity and their pairwise collision interaction is completely known. (In the case of the CH equation in 1D these solutions are solitons \cite{CaHo1993}.)

\paragraph{Emergent singularities in aggregation dynamics.}In recent years, another class of {\it dissipative} continuum models has been found to possess emergent singular solutions \cite{HP2005, HP2006}. These equations appeared in the study of aggregation and self-assembly phenomena at different scales in physics and biology.
The main idea is that the approach to a critical point in free energy of a continuum material may concentrate matter into self-organized structures (clumps).   Diverse examples of such processes include the formation of stars, galaxies and solar systems at large scales, growth of colonies of organisms at mesoscales and self-assembly of proteins, nanotubes or micro/nanodevices at micro- and nanoscales  \cite{Whitesides2002}. Some of these processes, such as  nano-scale self-assembly of molecules, are of great technological interest. 
\rem{ 
See \cite{Rabani2003,XB2004} for relevant experiments and a class of stochastic models for this process.  
Due to the large number of particles involved in nano-scale self-assembly ($10^9-10^{12}$), the development of continuum descriptions for aggregation or self-assembly is a natural approach
toward its theoretical understanding and modeling.
} 
  \smallskip

Mathematically, the classic examples of continuous equations for aggregation are those  of Debye-H\"uckel \cite{DeHu1923} and  Keller-Segel (KS) \cite{KellerSegel1970} for the density of particles. For a recent review of developments in this area with an emphasis on biophysical modeling, see, for example \cite{Ho03}. When diffusion is negligible, the physics of these models consists of a conservation law: $\partial_t \rho + \mbox{div} \left(\rho\,\mu\,\nabla \Phi\right)=0$, in which the ``mobility'' $\mu$ may also depend on the density, as it happens for the collective potential $\Phi[\rho]$. This simple relation is known as ``Darcy's Law'' \cite{Darcy1856}. Previous investigations have extended Darcy Law to incorporate nonlocal, nonlinear and anisotropic effects in self-organization of aggregating particles of finite size \cite{HP-GOP-2007}.

Besides the spontaneous emergence of singular solutions, the equations for aggregation dynamics exhibit self-assembly through the interaction among their singularities after they have emerged. Indeed, the study in \cite{HP2005, HP2006} has shown how the
{\bfi singular solutions clump together} after their formation. This feature reflects the physics underlying the mathematical description and distinguishes this class of dissipative equations from the energy-conserving equations in fluid mechanics, such as Euler's equation or the EPDiff equation mentioned above.

\paragraph{Geometric gradient-flow equations.} The goal of this paper is to formulate a principle for deriving and analyzing evolution equations for continuum systems that may exhibit self-assembly under a flow which reflects the mathematical properties of Darcy's Law, although involving an appropriate physical quantity possessing {\it other} geometric properties than particle density. A useful concept for deriving a continuum description of macroscopic  pattern formation (e.g., aggregation) due to microscopic  processes is the notion of {\bfi order parameter}. Order parameters are continuum variables that describe macroscopic effects due to microscopic variations of the material structure. They take values in a vector space called the order parameter space that respects the underlying geometric structure of the microscopic variables. A familiar example is the description of the local directional asymmetries of nematic liquid crystal molecules by a spatially and temporally varying macroscopic continuum field of unsigned unit vectors called ``directors,Ó see, e.g., Chandrasekhar \cite{Ch1992}  and de Gennes and Prost \cite{deGePr1993}.

\smallskip
In order to establish macroscopic continuum models for aggregation phenomena of order parameters, the present treatment formulates such equations by generalizing the geometric structure underlying Darcy's law. This is a non-trivial step, since Darcy's Law possesses an ambiguity when written in a covariant form using the Lie derivative, as explained in the first part of the paper.
This ambiguity may be eliminated by requiring the existence of singular solutions in the geometric gradient-flow (GGF) equations for the order parameters. The requirement of singular solutions is physically justified by the same arguments that hold for Darcy's Law, that is recovering self-assembly of many-particle systems in terms of coherent structures. 
Self-assembly will have a specific physical meaning, depending on the nature of the particular order parameter. For example, it will represent alignment-like phenomena in the case of rods or directors.

\paragraph{Plan of the work.} Section \ref{sec:DLaw} presents the geometric structure of Darcy's Law and the related ambiguity in its underlying geometric structure. Section~\ref{sec:GOP-eqn} identifies a unique geometric class of order parameter equations by requiring that singular solutions exist, so as to capture coherent structures by introducing a spatial averaging that follows them in a Lagrangian sense. 
In Section~\ref{Meth-MSV} this Lagrangian averaging approach yields nonlocal equations expressed in fixed spatial coordinates. 
Rather general mathematical and physical requirements (correct geometry of the measured variable and independence of the method of measurement) in Section~\ref{sec:thermodynamics} yield again the allowable equations discussed in Section ~\ref{sec:DLaw}.
Remarkably, these general considerations yield the same evolution equation (\ref{GOP-eqn-brkt}) for an arbitrary geometric quantity as found by using the Lagrangian averaging method. The ambiguity in the roles of the order parameter and its mobility is overcome by focusing on singular solutions resembling aggregation phenomena.
\rem{ 
, except for an ambiguity in the roles of the order parameter and its mobility, which live in the same vector space. We shall consider aggregation of geometric order parameters in continua that are either stationary, or flowing.
 \smallskip
 
In Sec.~\ref{sec:euler} we exploit the main properties of GGF equations and
their geometric structure. 
} 

In Section~\ref{sec:euler} we develop a form of dissipation in terms
of Leibnitz-bracket structures for for fluid mechanics, which reflect the
 Darcy dissipation term. As we show, this formalism leads to an interesting equation for modified dissipative fluid dynamics that does not require extra boundary conditions (unlike viscosity dissipation in Navier-Stokes equations). The formalism generalizes earlier modified fluid equations of this type in Bloch et al. \cite{BlKrMaRa1996, BlBrCr1997} and Vallis et al. \cite{VaCaYo1989}.
\smallskip

Next, in Section ~\ref{sec:scalars} we consider a self-organizing system of an arbitrary geometric quantity. 
The principle for obtaining such evolution equations applies generally for any geometric type of continuum physical quantity (density, scalar, 1-form, 2-form, {\it etc.}). Thus, we obtain a family of evolution equations for physical quantities such as active scalars, momenta and fluxes (Sec.~\ref{sec:1forms}) that nonlinearly influence their own evolution. These quantities convect themselves by inducing a velocity appropriate for continuum motion of any geometric order parameter. 
In other words, they can be written as nonlocal characteristic equations, as shown in 
Section ~\ref{sec:scalars}. 

After the geometric gradient-flow (GGF) equation  (\ref{GOP-eqn-brkt}) has been derived from general principles, we discuss its most remarkable feature; namely, this equation possesses \emph{singular} solutions. In these singular solutions, each type of order parameter may be localized into delta functions distributed along embedded subspaces moving through the ambient space. Depending on the geometric type of the order parameter, the space of singular solutions may either form an invariant manifold, or these solutions may emerge from smooth confined initial conditions. In the latter case, the singular solutions dominate the long-term aggregation dynamics. From the physical point of view, such localized, or quenched solutions would form the core of the processes of self-assembly and are therefore of great practical interest. The formation of these localized solutions is driven by a combination of  nonlinearity and nonlocality. Their evolution admits a reduced description, expressed completely in terms of coordinates on their singular embedded subspaces.

\section{Geometric gradient-flow equations}
\subsection{Geometric structure of Darcy law} \label{sec:DLaw}
The Darcy law for the geometric order parameter $\rho$
(density) \cite{HP2005,HP2006} is written in terms of an energy functional
$E=E[\rho]$ and a {\it mobility} $\mu$ which takes into account of the typical
size of the particles in the system (in general it depends on $\rho$). In
formulas, one has the equation
\begin{equation}\label{DLaw}
\frac{\pa \rho}{\pa t}=
{\rm div}\!\left(\rho\,\mu[\rho]\,\nabla\frac{\delta E}{\delta
\rho}\right)
\,.
\end{equation}
This may be stated by using the Lie derivative operator in the two possible ways,
\begin{equation}\label{exchange-darcy}
\text{either}\qquad
\frac{\pa \rho}{\pa t}=
\textit{\large\pounds}_\text{\!\footnotesize$\left(\rho\nabla\frac{\delta E}{\delta
\rho}\right)^{\!\sharp}$}\,\,\mu[\rho]
\qquad\text{or}\qquad
\frac{\pa \rho}{\pa t}=
\textit{\large\pounds}_\text{\!\footnotesize$\left(\mu[\rho]\nabla\frac{\delta E}{\delta \rho}\right)^{\!\sharp}$}\,\rho
\,,
\end{equation}
\rem{ 
\begin{quote}
``The local value of $\rho$ remains invariant along the characteristic curves of a flow, whose velocity $\bf u[\rho]$ depends on $\rho$.''
\end{quote}
This principle may be formulated in symbols as, 
\begin{eqnarray}
\frac{d\rho}{dt}(\mathbf{x}(t),t)=0
\quad\hbox{along}\quad
\frac{d\mathbf{x}}{dt}=\mathbf{u}[\rho]
\,.
\label{GOPprincip}
\end{eqnarray}
where the flow velocity $\mathbf{u}[\rho]$ is to be expressed .
In the case of particle density $\rho$ in $n$-dimensional space, for example, the number of particles
$\kappa=\rho\,\mbox{d}^n\mathbf{x}$ has physical meaning.
} 
where sharp $(\,\cdot\,)^\sharp$ denotes raising the vector index
from covariant to contravariant, so its divergence may be taken (the sign
in the right hand side is taken in agreement with the dissipative nature of the dynamics, as it is shown in Sec.~\ref{sec:geom-phys}).
The evident difference between these two forms is that, unlike the first
form, the second equation can be written as the characteristic equation 
\begin{eqnarray}
\frac{d\rho}{dt}(\mathbf{x}(t),t)=0
\quad\hbox{along}\quad
\frac{d\mathbf{x}}{dt}=\mathbf{u}[\rho]=
\left(\mu[\rho]\nabla\frac{\delta E}{\delta \rho}\right)^{\!\sharp}
\label{GOPprincip}
\end{eqnarray}
so that velocity $\mathbf{u}$ depends on density $\rho$ through the
gradient of the variation of the free energy
$E$ (velocity proportional to thermodynamic force with mobility $\mu[\rho]$)
\cite{HP2005,HP2006,HP-GOP-2007}.
\rem{ 
Here, the time
derivative of $\rho$ invokes the fundamental chain rule for the
product of the density function times the volume element,
$\rho(\mathbf{x}(t),t)\,\mbox{d}^n\mathbf{x}(t)$. Preservation of this
product along
$d\mathbf{x}/dt=\mathbf{u}[\rho]$ yields 
\begin{eqnarray}
(\partial_t \rho
+
\mathbf{u}[\rho]\cdot \nabla \rho
+ \rho\,{\rm div}\,\mathbf{u}[\rho] )\,\mbox{d}^n\mathbf{x}(t)
= 0
\,,
\end{eqnarray} 
As mentioned above, the Darcy Law approach assumes that velocity $\mathbf{u}$ depends on density $\rho$ through the
gradient of the variation of free energy
$E$ (velocity proportional to thermodynamic force with mobility $\mu[\rho]$)
\cite{HP2005,HP2006,HP-GOP-2007}. This assumption leads
to the expected continuity equation for density,
\begin{equation} 
\label{cont-eqn}
\partial_t \rho
=-\, {\rm div}\, 
\rho\,\mathbf{u}[\rho]
\quad\hbox{where}\quad
\mathbf{u}[\rho] = - (\mu\, \nabla \delta E/\delta \rho)^\sharp
\,,
\end{equation} 
and sharp $(\,\cdot\,)^\sharp$ denotes raising the vector index
from covariant to contravariant, so its divergence may be taken (the sign
in the expression of $\bf u$ is taken to be negative in agreement with the
dissipative nature of the dynamics, as it is shown in Sec.~\ref{sec:geom-phys}).
} 
\smallskip

As mentioned in the introduction, we want to generalize this geometric flow method underlying Darcy Law to apply to other order parameters (denoted by $\kappa$) with different geometrical meaning (not just densities).
\rem{ 
 Geometrical considerations might lead one to expect  that such a generalization would take the mathematical form,
\begin{eqnarray}
\partial\kappa/\partial t + \pounds_{u[\kappa]}\kappa=0
\,,
\label{GOPprincipmath}
\end{eqnarray}
where $\pounds_u\kappa$ denotes the Lie derivative with respect to the vector field $u=\mathbf{u}\cdot\nabla$ of any geometrical quantity $\kappa$ \cite{MaRa99}.
} 
The key question for understanding the physical modeling that would be needed in making such a generalization is, ``What is the corresponding Darcy Law for an order parameter $\kappa$?'' Namely, how does one determine the corresponding geometric flow for an arbitrary geometrical quantity $\kappa$? The first problem is that there is no reason to consider only one of the two geometric formulations in (\ref{exchange-darcy}). Although a characteristic form would be preferable because of its richer geometric  meaning, no choice can be performed a priori.

As a further step in the investigation of the geometric structure in Darcy
law (\ref{DLaw}), one wants to write the corresponding equation (\ref{DLaw})
in a fully covariant fashion. To this purpose, one defines the {\bfi diamond operator}
$\diamond$ as the \emph{dual} of the Lie derivative under integration by parts for any pair $(\kappa,b)$ of dual variables and any vector field $\mathbf{v}$ \cite{HP-GOP-2007}. That is 
\begin{eqnarray}
\langle \kappa \diamond b, \mathbf{v} \rangle=\langle \kappa, 
-\pounds_\mathbf{v\,} b \rangle
\,.
\label{diamond-def}
\end{eqnarray}
In this way the diamond operation is defined as $\diamond:V\times V^*\to\mathfrak{X}^*$,
where $V$ is the vector space which $\kappa$ belongs to and $V^*$ is its
dual, while $\mathfrak{X}^*$ denotes the dual space of vector fields in $\mathbb{R}^3$.
In the case of Darcy Law, $\kappa=\rho$ is a density variable and $b$ is
a scalar function, so that one has
\[
\langle \rho\diamond b,{\bf v} \rangle
=
-\,
\langle \rho,{\bf v}\cdot\nabla b \rangle
=
-\,
\langle \rho\nabla b,{\bf v} \rangle
\,.
\]
Thus, by using the definition of diamond, one expresses Darcy Law (\ref{DLaw}) in covariant form as
\begin{equation}\label{exchange-diamond}
\text{either}\qquad
\frac{\pa \rho}{\pa t}+
\textit{\large\pounds}_\text{\!\footnotesize$\left(\rho\diamond\frac{\delta E}{\delta
\rho}\right)^{\!\sharp}$}\,\,\mu[\rho]=0
\qquad\text{or}\qquad
\frac{\pa \rho}{\pa t}+
\textit{\large\pounds}_\text{\!\footnotesize$\left(\mu[\rho]\diamond\frac{\delta E}{\delta \rho}\right)^{\!\sharp}$}\,\rho=0
\,.
\end{equation}
This formulation is not only covariant, but it can as well be extended to
any quantity for which a Lie derivative can be defined, i.e. any tensor field, or {\bfi geometric order parameter} $\kappa$. It only suffices to substitute $\rho$ with $\kappa$ in order to obtain the following two possibilities:
\begin{equation}\label{kappa-exchange}
\text{either}\qquad
\frac{\pa \kappa}{\pa t}=-\,
\textit{\large\pounds}_\text{\!\footnotesize$\left(\kappa\diamond\frac{\delta E}{\delta
\kappa}\right)^{\!\sharp}$}\,\,\mu[\kappa]
\qquad\text{or}\qquad
\frac{\pa \kappa}{\pa t}=-\,
\textit{\large\pounds}_\text{\!\footnotesize$\left(\mu[\kappa]\diamond\frac{\delta E}{\delta \kappa}\right)^{\!\sharp}$}\,\kappa
\,.
\end{equation}
It is important to notice that, unlike Darcy's Law, the equations above are {\it not} invariant with respect to the substitution $\kappa\leftrightarrow\mu[\kappa]$
and the two equations in (\ref{kappa-exchange}) are {\it different} equations.

\subsection{GGF equations and singular solutions}\label{sec:GOP-eqn}
At this point it is necessary to make a definitive choice between these two possibilities, although both appear to be consistent geometric extensions of Darcy Law (\ref{DLaw}). Actually, one recalls that equation (\ref{DLaw}) with generic mobility $\mu=\mu[\rho]$ has one more feature, besides its purely geometric character. This feature is the {\bfi spontaneous emergence of singular solutions}, that correspond in one dimension to the trajectories of $N$ interacting particles in the system \cite{HP2005,HP2006}. This point leads
to the question: is it possible to generalize the existence of singular solutions to geometric gradient-flow (GGF) equations? Thus one is motivated to look at one of the forms in (\ref{kappa-exchange})
and see whether they have this property. For example, in one dimension one
looks for solutions of the form
\[
\kappa(x,t)=\sum_{i=1}^N p_i(t)\,\delta(x-Q_i(t))
\,.
\]
Upon pairing the first equation in (\ref{kappa-exchange}) with a dual element $\phi$, direct substitution of the singular solution ansatz yields
\begin{multline}
\bigg\langle
\frac{\pa\kappa}{\pa t},\,\phi
\bigg\rangle 
=
\sum_i\,\frac{\partial p_i}{\partial t}\cdot \phi \left({Q}_i(t) \right)
+
\sum_i\,
\frac{\partial Q_i}{\partial t}\cdot \phi^{\,\prime}\left(Q_i(t) \right) 
\\
=-\,
\bigg\langle 
\textit{\large\pounds}_\text{\!\footnotesize$\left(\kappa\diamond\frac{\delta E}{\delta \kappa}\right)^{\!\sharp}$}\,\mu\,,\,\phi
\bigg\rangle
=-\,
\bigg\langle 
\mu\diamond\phi,\,
\left(\kappa\diamond\frac{\delta E}{\delta \kappa}\right)^{\!\sharp}
\bigg\rangle
=-\,
\bigg\langle 
\kappa,\,
\textit{\large\pounds}_\text{\!\small$\left(\mu\diamond\phi\right)^{\sharp}$}\,\text{\small$\frac{\delta E}{\delta \kappa}$}
\bigg\rangle
\\
=-\,
\sum_{i=1}^N\,
p_i(t)\contract
\left.
\textit{\large\pounds}_\text{\!\small$\left(\mu\diamond\phi\right)^{\sharp}$}\,\text{\small$\frac{\delta E}{\delta \kappa}$}\,
\right|_{x=Q_i(t)}
\end{multline}
where the symbol $\bf\contract$ denotes contraction of indexes.
In order for the singular solutions to exist, one would match terms in $\phi$
and $\phi^{\,\prime}$ and obtain the evolution equations for $p_n$ and $Q_n$,
as it happens for the density variable $\rho$ in Darcy law \cite{HP2005,HP2006}. However, in general the term in the last line may involve higher derivatives, not just first order (for example, if $\kappa$ is a one-form density, then
diamond is again a Lie derivative, which generates second order derivatives
in $\phi$). Therefore, the first choice in (\ref{kappa-exchange}) is not suitable to recover the singular solutions in the general case of an order parameter $\kappa$. Instead, by following the same procedure for
the second equation in (\ref{kappa-exchange}), one obtains
\begin{multline}
\label{calc-sing}
\bigg\langle
\frac{\pa\kappa}{\pa t},\,\phi
\bigg\rangle 
=
\sum_i\,\frac{\partial p_i}{\partial t}\cdot \phi \left({Q}_i(t) \right)
+
\sum_i\,
\frac{\partial Q_i}{\partial t}\cdot \phi^{\,\prime}\left(Q_i(t) \right)
=-\,
\sum_{i=1}^N\,
p_i(t)\contract
\left.
\textit{\large\pounds}_\text{\!\footnotesize$\left(\mu\diamond\frac{\delta E}{\delta
\kappa}\right)^{\sharp}$}\,\phi\,
\right|_{x=Q_i(t)}
\end{multline}
Now, since the Lie derivative is a first-order differential operator that is linear in the derivatives, one recognizes that the last term on the second line contains only terms that are {\it linear} in $\phi$ and its first order derivatives, while it does not involve any of its higher order derivatives. Thus, in higher dimensions one finds the following conclusion
\begin{theorem}
The second equation of (\ref{kappa-exchange}) always allows for singular
solutions of the form
\begin{equation}
\kappa({\bf x},t)=\sum_{i=1}^N \int p_i(s,t)\,\delta({\bf x-Q}_i(s,t))\,{\rm d}s
\label{GOPsing}
\end{equation}
for any tensor field $\kappa$, provided $\mu$ and $\delta E/\delta \rho$ are sufficiently smooth.
\end{theorem}
The proof of this statement is a verification through direct substitution,
 similar to the calculation in equation (\ref{calc-sing}). As in the introduction, the variable $s$ is a coordinate on a submanifold
of $\mathbb{R}^3$: if $s$ is a one-dimensional coordinate, then $\kappa$
is supported on a curve (filament), if $s$ is two dimensional, then $\kappa$
is supported on a surface (sheet) immersed in physical space. 

At this point, one defines {\bfi geometric gradient-flow (GGF) equations} as characteristic equations of the
type
\begin{eqnarray}
\frac{d\kappa}{dt}(\mathbf{x}(t),t)=0
\quad\hbox{along}\quad
\frac{d\mathbf{x}}{dt}=
\left(\mu[\kappa]\nabla\frac{\delta E}{\delta \kappa}\right)^{\!\sharp}
\label{GOP-char}
\end{eqnarray}
or, in Eulerian coordinates,
\begin{equation}
\frac{\pa \kappa}{\pa t}=-\,
\textit{\large\pounds}_\text{\!\footnotesize$\left(\mu[\kappa]\diamond\frac{\delta E}{\delta \kappa}\right)^{\!\sharp}$}\,\kappa
\label{HP-kappa}
\end{equation}
where the quantity $(\mu\diamond\delta E/\delta\kappa)^\sharp$ can be called
{\it generalized Darcy velocity} and in the case of Darcy Law expresses the
fact that the macroscopic velocity is assumed proportional to the collective force
${\bf v}\propto\nabla\Phi$, where the coefficient of proportionality is given
by the mobility $\mu$. The fact that Darcy law (\ref{DLaw}) is invariant
with respect to permutations of $\rho$ and $\mu$ is the reason why singular solutions are recovered for both possibilities in (\ref{exchange-diamond}). This property is peculiar of Darcy law and does not hold in general. The distinction between the two cases identifies the geometric structure of the GGF family of equations.

\paragraph{Variational formulation.}
As a further step one seeks a variational formulation. One defines a gradient flow by setting the $L^2$ pairing of equation (\ref{HP-kappa}) with a test variable $\phi$ (dual to $\kappa$) equal to the variation $\delta E$ of the free energy
\begin{eqnarray}
\label{gradflow-def}
\left\langle \frac{\pa \kappa}{\pa t} \,,\, {\phi} \right\rangle 
=
\left\langle 
\delta\kappa \,,\, 
\frac{\delta E}{\delta \kappa} 
\right\rangle 
\,,
\end{eqnarray}
where the variation $\delta \kappa$ satisfies
\begin{eqnarray}
\label{geomvar-def}
\delta \kappa
=
-\,\pounds_\text{\footnotesize$\big(\kappa\diamond \phi \big)^\sharp$}\,\mu[\kappa]
\,,
\end{eqnarray}
in order to recover equation (\ref{HP-kappa}). Indeed, one
obtains
\begin{align}
\Big\langle \frac{\partial \kappa}{\partial t} \,,\, {\phi} \Big\rangle 
=
\bigg\langle 
\delta \kappa\,,\, 
\frac{\delta E}{\delta \kappa} 
\bigg\rangle 
=
-\,
\bigg\langle 
\pounds_\text{\footnotesize$\big(\kappa\diamond \phi \big)^\sharp$}\,\mu \,,\, 
\frac{\delta E}{\delta \kappa} 
\bigg\rangle 
&=:
-\,
\bigg\langle 
\mu \diamond
\frac{\delta E}{\delta \kappa} \,,\,
\big(\kappa\diamond \phi \big)^\sharp
\bigg\rangle 
\label{HP-rho-calc}
\,.
\end{align}

The next question in the formulation of GGF theory is the particular meaning assumed by the {\it generalized mobility} $\mu[\kappa]$. This quantity has been related to the typical particle size in Darcy dynamics \cite{HP2005,HP2006}, but the physical meaning of this quantity is not yet clear in the case of a generic GGF equation for the order parameter $\kappa$. The next section
presents the mobility as a smoothed quantity that keeps into account the dynamics of jammed states in the system, by introducing a typical length-scale.

\subsection{Motivation for multi-scale variations}\label{Meth-MSV}
We seek a variational principle for a continuum description of coherent structures. This includes the evolution of particles of finite size that may clump together under crowded conditions. In crowded conditions, finite-sized particles typically reach jammed states, sometimes called  rafts, that may be locally locked together over a coherence length of several particle-size scales. Thus, a variational principle for the evolution of coherent structures such as jammed states in particle aggregation must accommodate more than one length scale. A  multi-scale variational principle may be derived by considering the variations as being applied to rafts, or patches, of jammed states of a certain size (the coherence length). For example, one might introduce a probability distribution for the sizes formed in the process of particle clumping. We shall take a simpler approach based on applying a Lagrangian coordinate average that moves with the clumps of particles. In this approach, the variation ($\delta\kappa$) of the order parameter ($\kappa$) at a given fixed point in space is determined by a family of smooth maps $\varphi(s)$ depending continuously on a parameter $s$ and acting on the average value $\bar{\kappa}$ defined by
\begin{eqnarray}
\bar{\kappa} = H*\kappa = \int H(y-y')\kappa(y') dy'
\end{eqnarray}
in the frame of motion of the jammed state. Thus, $y$ is a Lagrangian coordinate, defined in that frame. This motion itself is to be determined by the variational principle. The kernel $H$ represents the typical size and shape of the coherent structures, which in this example would be rafts of close-packed finite-size particles. The mobility $\mu$ of the rafts depends on $\bar{\kappa}$, and so the corresponding variation of the local quantity $\kappa$ at a fixed point in space may be modeled as
\begin{eqnarray}
\delta\kappa 
=\frac{d}{ds}\Big|_{s=0}\Big(\mu(\bar{\kappa})\varphi^{-1}(s)\Big)
\,.
\end{eqnarray}
Here $\varphi(s)y=x(s)$ is a point in space, which $\varphi^{-1}(s)$ returns  to its Lagrangian label $y$ and $\varphi(0)$ is the identity operation.
The average $\bar{\kappa}=H*\kappa$ is applied in a \emph{Lagrangian} sense, following a locally locked raft of particles along a curve parameterized by time $t$  in the family of smooth maps. The latter represents the motion of the raft as  $\varphi(t)y=x(t)$, whose velocity tangent vector is still to be determined. When composed from the right the derivative at the identity of the action of $\varphi(s)$ results in a variation $\delta\kappa$ at a fixed point in space given by
\begin{eqnarray}
\delta\kappa =-\pounds_{\mathbf{v}(\varphi)}\mu[\kappa]
\quad\hbox{with}\quad
\mathbf{v}(\varphi) = \varphi'\varphi^{-1}\big|_{s=0}
\,,
\end{eqnarray}
where $\pounds_{\mathbf{v}(\varphi)}\mu[\kappa]$ denotes the Lie derivative of the mobility $\mu[\kappa]$ with respect to the vector field $\mathbf{v}(\varphi)$.
In what follows we will treat $\mu[\kappa]$ as a general functional of $\kappa$, not just a function of $\bar{\kappa}$. \smallskip

Taking the steps for $\kappa$ corresponding to those for $\rho$ in (\ref{HP-rho-calc}) but with the Lie derivative appropriate for the geometric nature of $\kappa$ allows one to close the GGF equation (\ref{HP-kappa}) explicitly as
\begin{eqnarray}
\label{GOP-eqn-brkt}
\frac{\partial \kappa}{\partial t}
=
-\,\pounds_{{u}[\kappa]}
\kappa
\,,
\quad\hbox{in which}\quad
{u}[\kappa]
=
\Big(\mu[\kappa]\, {\diamond}\frac{\delta E}{\delta \kappa} \Big)^\sharp
\,.
\end{eqnarray}
Here the co-vector $(\mu[\kappa]\,\diamond\delta E/\delta \kappa)$ resulting from the diamond operation is defined as in equation (\ref{diamond-def}) and sharp $(\,\cdot\,)^\sharp$ raises its vector index to make it contravariant. Although Lagrangian averaging was invoked in defining the mobility as a coherence property, equation 
(\ref{GOP-eqn-brkt}) is expressed in fixed spatial coordinates. Its Lagrangian heritage is recognized upon rewriting it equivalently in the characteristic form (\ref{GOPprincip}).

\subsection{A dissipative bracket for GGF equations }
\label{sec:geom-phys}
We shall use the method of multi-scale variations to derive several
versions of the geometric gradient-flow (GGF) equation (\ref{HP-kappa}) for different geometrical types of physical quantities $\kappa$. In each case, the result is a novel nonlocal characteristic equation which possesses localized (singular) solutions. Before developing these results, however, we shall first address the question of energy dissipation. This will allow us to introduce the {\bfi dissipative bracket} and to write the equations in an alternative bracket form. 
The corresponding energy equation
follows from (\ref{HP-kappa}) as
\begin{eqnarray}
\frac{d E}{dt}
=
\Big\langle \frac{\partial \kappa}{\partial t} \,,\, 
\frac{\delta E}{\delta \kappa} \Big\rangle
&=&
\left\langle  -\,
\pounds_{(\mu[\kappa]\, {\diamond}\frac{\delta E}{\delta \kappa})^\sharp}
\kappa,\,\frac{\delta E}{\delta \kappa}
\right\rangle
\, 
=
\, 
-\,
\left\langle  
\Big(\mu[\kappa] \,\diamond\, \frac{\delta E}{\delta \kappa}\Big)
,\, 
\Big(\kappa\,\diamond\,\frac{\delta E}{\delta \kappa}\,\Big)^\sharp
\right\rangle
.
\label{GOP-erg}
\end{eqnarray}
The formula for energy in (\ref{GOP-erg}) suggests the following bracket notation for the time derivative of a functional $F[\kappa]$,
\begin{eqnarray}
\frac{d F[\kappa]}{dt}
=
\Big\langle \frac{\partial \kappa}{\partial t} \,,\, 
\frac{\delta F}{\delta \kappa} \Big\rangle
&=&
\left\langle  
-\,\pounds_{(\mu[\kappa]\, {\diamond}\frac{\delta E}{\delta \kappa})^\sharp}
\kappa
\,,\,
\frac{\delta F}{\delta \kappa}
\right\rangle
\nonumber \\
&=&
-\,
\left\langle  
\Big(\mu[\kappa] \,\diamond\, \frac{\delta E}{\delta \kappa}\Big)
,\, 
\Big(\kappa\,\diamond\,\frac{\delta F}{\delta \kappa}\,\Big)^\sharp
\right\rangle
=:
\{\!\{\,E\,,\,F\,\}\!\} [\kappa]
\label{bracketdef}
\end{eqnarray}
The properties of the GGF brackets $ \{\!\{\,E\,,\,F\,\}\!\}$ defined in equation (\ref{bracketdef}) are determined by the diamond operation and the choice of the mobility $\mu[\kappa]$. For physical applications, one should choose a mobility that satisfies strict dissipation of energy, \emph{i.e.}
$
  \{\!\{\,E\,,\,E\,\}\!\} \leq 0. 
$
A particular example of mobility that satisfies the energy dissipation requirement  is $\mu[\kappa]=\kappa M[\kappa]$, where $M[\kappa] \geq 0$ is a non-negative scalar functional of $\kappa$. (That is, $M[\kappa]$ is a number.) Requiring the mobility to produce energy dissipation does not limit the mathematical properties of the GGF bracket.  For example, the dissipative bracket possesses the Leibnitz property with any choice of mobility. That is, it satisfies the Leibnitz rule for the derivative of a product of functionals. In addition, the dissipative bracket formulation (\ref{bracketdef}) allows one to reformulate the GGF equation (\ref{HP-kappa}) in terms of flow on a Riemannian manifold with a metric defined through the dissipation bracket, as discussed in more detail in \cite{HP-GOP-2007}. 

\paragraph{Previous dissipative brackets.} Historically, the use of symmetric brackets for introducing dissipation into Hamiltonian systems seems to have originated with works of Grmela \cite{Gr84}, Kaufman \cite{Ka1984} and Morrison \cite{Mo1984}. See \cite{Ot05} for references and further engineering developments. This approach introduces a sum of two brackets, one describing the Hamiltonian part of the motion and the other obtained by representing the dissipation with a symmetric bracket operation involving an entropy defined for that purpose. 
Being expressed in terms of the diamond operation $(\,\diamond\,)$ for an arbitrary geometric order parameter $\kappa$, the dissipative bracket in equation (\ref{bracketdef}) differs from symmetric brackets proposed in earlier work.  The  geometric advection law (\ref{GOP-eqn-brkt}) for the order parameter will be shown below to arise from thermodynamic principles that naturally yield the dissipative bracket (\ref{bracketdef}).  Moreover, being written as a Lie derivative, the equation of motion (\ref{GOP-eqn-brkt}) respects the geometry of the transported quantity. The dissipative brackets from the earlier literature do not appear to be expressible as a geometric transport equation in Lie derivative form. 

\subsection{Thermodynamic and geometric implications
}
\label{sec:thermodynamics} 
Equations (\ref{GOP-eqn-brkt}) may be justified using general principles of thermodynamics and geometry. Consider using an arbitrary functional $F$  in (\ref{bracketdef}) as a basis for the derivation of an equation for $\kappa$. 
Suppose $\kappa$ is an observable quantity for a physical system, and that system evolves due to the inherent free energy $E[\kappa]$ in the absence of external forces. This is the physical picture one envisions, for example, when thinking about any process of aggregation. Suppose we would like to measure the time evolution of a functional $F[\kappa]$, which may for example represent total energy or total momentum. For an \emph{arbitrary} functional $F[\kappa]$ and for a  \emph{given} free energy $E[\kappa]$ one finds, in general 
\begin{equation} 
\frac{d F}{d t}
=
 \bigg\langle 
 \partial_t \kappa \, , \, \frac{\delta F}{\delta \kappa}
 \bigg \rangle 
=
 \bigg\langle 
 \delta \kappa \, , \, \frac{\delta E}{\delta \kappa}
 \bigg \rangle 
 =:
 \delta E\Big|_F
 \,,
 \label{arbitraryF}
\end{equation} 
where the variation $\delta \kappa$ is defined in terms of the functional $F$ as in equation (\ref{geomvar-def}), namely,
\begin{eqnarray}
\label{geomvar-F}
\delta \kappa
=
-\,\pounds_\text{\footnotesize$\big(\kappa\diamond \delta F/\delta \kappa \big)^\sharp$}\,\mu[\kappa]
\,.
\end{eqnarray}
That is, all steady states (for the evolution of any functional $F$) will be critical points of the free energy with respect to variations in free energy determined from the definition of $F$ by equation (\ref{geomvar-F}) in terms of $\mu[\kappa]$.

The main thermodynamic postulate  here is that, in principle, one can determine the evolution of the system by probing in many different directions associated with \emph{global} quantities $F[\kappa]$ (for example, the moments of a probability distribution). It is only natural to assume that the law for the evolution of $\kappa$ should be independent of the choice of which quantities $F$ are used to determine it. 
Surprisingly, this rather general sounding assumption sets severe restrictions on the nature of the variation $\delta \kappa$. In particular, 
\vspace{-3mm}
\begin{enumerate} \itemsep -2mm
\item The variation $\delta \kappa$ must be linear in $\delta F/\delta \kappa$, since the left hand side of (\ref{arbitraryF}) is also linear in $\delta F/\delta \kappa$. 
\item The variation $\delta \kappa$ must transform the same way as $\kappa$, as it must be dual to $\delta F/\delta \kappa$. This introduces the \emph{mobility}  $\mu$ that must be of the same type as $\kappa$. 
\item  The variation $\delta \kappa$ must specify a quantity at the tangent space to the space of all possible $ \kappa$. The proper geometric way to specify this quantity is through the Lie derivative $\pounds_\mathbf{v}$ with respect to some vector field $\mathbf{v}$.  
\end{enumerate} \vspace{-3mm}

There are only two ways to specify $\delta \kappa$ so that it  obeys these three thermodynamic and geometric constraints, when we insist that only a single new physical quantity $\mu[\kappa]$ is introduced. Namely, 
\begin{eqnarray} 
\label{choice1}
 \mbox{either} \quad 
\delta \kappa = - \pounds_{\mathbf{v}} \kappa  
& \quad  \mbox{with} \quad 
& 
\mathbf{v}=\left( \mu[\kappa] \diamond \frac{\delta F}{\delta \kappa} \right)^\sharp
\,, 
 \\  \mbox{or} \quad
\delta \kappa = - \pounds_{\mathbf{v}} \mu[\kappa] 
& \quad \mbox
{with} \quad 
& 
\mathbf{v}= \left( \kappa \diamond \frac{\delta F}{\delta \kappa} \right)^\sharp
\,.
\label{choice2} 
\end{eqnarray} 
Both of (\ref{choice1}) and (\ref{choice2}) are consistent with all three geometric and thermodynamics requirements. However, the first possibility (\ref{choice1}) prevents formation of measure-valued solutions in $\kappa$, when $\kappa$ is chosen to be a 
1-form, a 2-form or a vector field, and we know that these solutions must exist.  In contrast, the second possibility (\ref{choice2}) leads to (\ref{GOP-eqn-brkt}), which does admit measure-valued solutions for an arbitrary geometric quantity $\kappa$. In the remainder of the paper we shall choose (\ref{choice2}) and investigate the corresponding evolution equation (\ref{GOP-eqn-brkt}). The alternative choice (\ref{choice1}) would have reversed the roles of $\kappa$ and $ \mu[\kappa] $ in the Lie derivative.\\

\section{GGF  vortex dynamics}
\label{sec:euler}
An important physical system in the context of the geometric dynamics of exact two-forms is the Lie-Poisson system for the vorticity of an ideal Euler fluid \cite{MaWe83}. 
The developments above produce an interesting opportunity for the addition of dissipation to ideal fluid equations. This opportunity arises from noticing that the dissipative diamond flows that were just derived could just as well be used with any type of evolution operator, not just the Eulerian partial time derivative. 
For example, if we choose the geometric order parameter $\kappa$ to be the exact two-form $\omega=\bom\cdot d\mathbf{S}$ appearing as the vorticity in Euler's equations for incompressible motion with fluid velocity $\mathbf{u}$, then the GGF  equation (\ref{GOP-eqn-brkt}) with Lagrangian time derivative may be introduced as a modification of Euler's vorticity equation as follows,
\begin{equation}
\underbrace{\
\partial_t\,\omega+\pounds_u \omega\
}
_{\hbox{Euler}}
=
\underbrace{\
-\,\pounds_{(\mu[\omega]\, {\diamond}\frac{\delta E}{\delta \omega})^\sharp}
\omega\
}
_{\hbox{Dissipation}}
\,.
\label{vortdiss}
\end{equation}
Euler's vorticity equation is recovered when the left hand side of this equation is set equal to zero. This modified geometric form of vorticity dynamics supports point vortex solutions, requires no additional boundary conditions, and dissipates kinetic energy for the appropriate choices of $\mu$ and $E$.  Equation (\ref{vortdiss}) will be derived after making a few remarks about the geometry of the vorticity governed by Euler's equation. \smallskip

The Lie-Poisson bracket for vorticity is written on the dual  
$\mathfrak{X}^*_\textnormal{vol}$ of the Lie-algebra $\mathfrak{X}_\textnormal{vol}$ of volume-preserving diffeomorphisms, which is isomorphic to the set of exact one-forms $\omega=d\alpha$, where $\alpha$ is a generic one-form. In this case the Jacobi-Lie bracket between two volume-preserving vector fields $\boldsymbol{\xi}_1$ and $\boldsymbol{\xi}_2$ in $\mathbb{R}^3$ may be written as
\[
\left[\,\boldsymbol{\xi}_1\,,\,\boldsymbol{\xi}_2\,\right]\,
=-\,\text{curl}\left(\,\boldsymbol{\xi}_1\,\times\,\boldsymbol{\xi}_2\,\right)
\,.
\]
In terms of the vector potentials for which $\boldsymbol{\xi}_1=\text{curl }\boldsymbol{\psi}_1$ and $\boldsymbol{\xi}_2=\text{curl }\boldsymbol{\psi}_2$ this bracket becomes
\[
\left[\,\boldsymbol{\xi}_1\,,\,\boldsymbol{\xi}_2\,\right]\,
=-\,\text{curl }\big(\,\text{curl }\boldsymbol{\psi}_1\,\times\,\text{curl }\boldsymbol{\psi}_2\,\big)\,.
\]
The vector potentials $\boldsymbol{\psi}_1$ and $\boldsymbol{\psi}_2$ are defined up to a gradient of a scalar function
so that one can always choose a gauge in which $\text{div}\,\boldsymbol\psi=0$. Pairing the vector field given by the Lie bracket with a one-form (density) $\boldsymbol{\alpha}$ then yields,  after an integration by parts,
\[
\left\langle
\boldsymbol\alpha\,,\,\left[\,\boldsymbol{\xi}_1\,,\,\boldsymbol{\xi}_2\,\right]
\right\rangle
=
-\big\langle\,
\text{curl }\boldsymbol{\alpha}\,,\,\big(\,\text{curl }\boldsymbol{\psi}_1\,\times\,\text{curl }\boldsymbol{\psi}_2\,\big)
\,\big\rangle
=
-\big\langle\,
\text{curl }\boldsymbol{\alpha}\,,\,\big[\!\big[\,\boldsymbol{\psi}_1\,,\,\boldsymbol{\psi}_2\,\big]\!\big]
\,\big\rangle
\]
where we have defined
\[
\big[\!\big[\,\boldsymbol{\psi}_1\,,\,\boldsymbol{\psi}_2\,\big]\!\big]\,
:=\,
\text{curl }\boldsymbol{\psi}_1\,\times\,\text{curl }\boldsymbol{\psi}_2\,.
\]
The operation $[[\,\cdot\,,\,\cdot\,]]$ defines a Lie algebra structure on the space of vector potentials whose dual space may be naturally identified with exact two-forms $\omega=\text{curl}\,\alpha$. At this point, the expression for the Lie-Poisson bracket for functionals of vorticity may be introduced as
\[
\{F,\,H\}
\,=\,
\Big\langle
\bom\,,\, \Big[\! \Big[\,
\frac{\delta F}{\delta\bom}\,,\,\frac{\delta H}{\delta\bom}
\, \Big]\! \Big]
\Big\rangle
\,=\,
\int\boldsymbol\omega\,\cdot
\left(
\text{curl}\,\frac{\delta F}{\delta\bom}
\times
\text{curl}\,\frac{\delta H}{\delta\bom}
\right)\,\text{d}^3\bf x
\,,
\]
where $H=\frac{1}{2}\int \bom\cdot(-\Delta)^{-1}\bom\,d\,^3x$ is the fluid's kinetic energy expressed in terms of vorticity. 

Now,  vorticity dynamics is an example of geodesic motion on a
(infinite-dimensional) Lie group \cite{Ar1966}
\[
\partial_t\,\boldsymbol\omega
\,=\,
\{\bom,\,H\}
\,=\,
-\,{\rm ad}^*_{\,\delta H/\delta \boldsymbol\omega}\,\,\boldsymbol\omega\,=-\,
\text{curl}\big(\boldsymbol\omega\times\text{curl}\,(-\Delta)^{-1}\boldsymbol\omega\big)=
\text{curl}\big(\text{curl}\,\boldsymbol\psi\times\boldsymbol\omega\big)
=
-\,\pounds_{\,\text{curl}\boldsymbol\psi}\,\, \boldsymbol\omega
\]
and allows singular solutions in the form of vortex filaments.

In order to write the GGF evolution equation (\ref{GOP-eqn-brkt}) for $\boldsymbol\omega$ one
must compute the diamond operation $\diamond$ for the ${\rm ad}^*$ action, which is defined in terms of Lie derivative by
\begin{equation}
{\rm ad}^*_{\boldsymbol\psi}\bom
=
\pounds_{{\rm curl}\,\boldsymbol\psi}\,\bom
\,.
\end{equation}
The computation of the $\diamond$ operation follows from its definition in equation (\ref{diamond-def}). For any two velocity vector potentials $\boldsymbol\phi$ and $\boldsymbol\psi$, and an exact two form $\bom$ one finds
\begin{align}
\left\langle
\boldsymbol\phi\diamond\bom,\boldsymbol\psi
\right\rangle
&=
-\,
\left\langle
\boldsymbol\phi,\pounds_{{\rm curl}\,\boldsymbol\psi}\,\bom
\right\rangle
=
\left\langle
\boldsymbol\phi,\text{curl}\big(\bom\times\text{curl}\,\boldsymbol\psi\big)
\right\rangle\\
&=
\left\langle\,
\text{curl}\,\boldsymbol\phi\times\bom,\text{curl}\,\boldsymbol\psi\,
\right\rangle
=
\left\langle\,
\text{curl}\,\big(\text{curl}\,\boldsymbol\phi\times\bom\big),\boldsymbol\psi\,
\right\rangle
\,.
\end{align}
Consequently, up to addition of a gradient, the diamond operation is given in vector form as
\begin{equation}
\boldsymbol\phi\diamond\bom=
\text{curl}\,\big(\text{curl}\,\boldsymbol\phi\times\bom\big)
\,.
\label{diamond-vectot-pot}
\end{equation}
The insertion of this expression in the bracket (\ref{bracketdef}) gives the GGF equation for $\bom$,
\begin{equation}
\partial_t\,\boldsymbol\omega
=
\text{curl}\left(\boldsymbol\omega\times\text{curl curl}\left(\boldsymbol\mu[\boldsymbol\omega]\times
\text{curl}\,\frac{\delta E}{\delta\boldsymbol\omega}\,\right)\right)
\,.
\label{GOP-eqn-omega}
\end{equation}
Consequently, equation (\ref{vortdiss}) emerges in the equivalent forms, 
\begin{align}
\partial_t\,\boldsymbol\omega
&=
-\,{\rm ad}^*_{\boldsymbol\psi}\, \boldsymbol\omega\,
+\,{\rm ad}^*_{\,\,({\rm ad}^*_{\boldsymbol\psi}\,\, \boldsymbol\mu[\boldsymbol\omega])^{\,\sharp}}\,\, \boldsymbol\omega
\\
&=
\text{curl}\left(\boldsymbol\omega
\times\text{curl}\left(\,-\,\frac{\delta H}{\delta\boldsymbol\omega}
+ \text{curl}\left(\boldsymbol\mu[\boldsymbol\omega]
\times
\text{curl}\,\frac{\delta E}{\delta\boldsymbol\omega}\right)
\,\right)\right)
\,.
\label{vortdiss1}
\end{align}
The full dynamics for the vorticity in equation (\ref{vortdiss}) is specified up to the choices of the mobility $\boldsymbol\mu[\bom]$ and the energy in the dissipative bracket $E[\bom]$. By definition, the mobility belongs to the dual space of volume-preserving
vector fields which is here identified with exact two-forms, thus one can
write the mobility in terms of its vector potential as $\boldsymbol\mu=\text{curl}\,\boldsymbol\lambda$ and rewrite the GGF equation (\ref{GOP-eqn-omega}) as 
\begin{equation}
\partial_t\,\bom
=
\text{curl}\Big(\bom\times\text{curl}\,\text{curl}
\Big[\! \Big[\,\boldsymbol\lambda\,,\,\frac{\delta E}{\delta\bom}\, \Big]\! \Big] \Big)
\,.
\end{equation}

This equation raises questions concerning the dynamics
of vortex filaments with nonlocal dissipation, following the ideas in \cite{Ho03}, where connections were established between the Marsden-Weinstein bracket \cite{MaWe83} and the Rasetti-Regge bracket for vortex dynamics \cite{RaRe, Ho2003}. Ideas for dissipative bracket descriptions in fluids have been introduced previously, see Bloch et al. \cite{BlKrMaRa1996, BlBrCr1997} and references therein. In particular, equation (\ref{vortdiss1}) recovers equations (2.2-2.3) of Vallis et al. \cite{VaCaYo1989} when $E=H$ and $\boldsymbol\mu=\alpha\bom$ for a constant $\alpha$. 

\begin{remark}
The GGF equation (\ref{vortdiss}) may be expressed as 
the Lie-derivative relation for conservation of vorticity flux, 
\begin{equation}
\partial_t\,(\bom\cdot d\mathbf{S})
\,=\,-\,\pounds_{\mathbf{u}+\mathbf{v}}\,
(\bom\cdot d\mathbf{S})
\,,
\label{Lie-der-GOP-vortex}
\end{equation}
in which the velocities $\mathbf{u}$ and $\mathbf{v}$ may be written in terms of the commutator $[\,\cdot\,,\,\cdot\,]$ of divergenceless vector fields as,
\begin{equation}
\mathbf{u}
=
{\rm curl}\,\frac{\delta H}{\delta\boldsymbol\omega}
\,,\quad
\mathbf{v}
=-\,
{\rm curl}\,{\rm curl}\left(\boldsymbol\mu[\boldsymbol\omega]
\times
\mathbf{\tilde{u}} \right)
=
{\rm curl}\,\big[\,\boldsymbol\mu[\bom]
\,,\,\mathbf{\tilde{u}}\,\big]
\quad\hbox{where}\quad
\mathbf{\tilde{u}}
=
{\rm curl}\,\frac{\delta E}{\delta\boldsymbol\omega}
\,,
\label{uandv-3D}
\end{equation}
Since both $\mathbf{u}$ and $\mathbf{v}$ are divergenceless, the vorticity equation (\ref{Lie-der-GOP-vortex}) may also be expressed as a commutator of divergenceless vector fields, denoted as $[\,\cdot\,,\,\cdot\,]$,
\begin{equation}
\partial_t\,\bom + (\mathbf{u}+\mathbf{v})\cdot\nabla\bom
-
\bom\cdot\nabla(\mathbf{u}+\mathbf{v})
=
\partial_t\,\bom
+
[\mathbf{u}+\mathbf{v},\,\bom]
=
0
\,.
\label{vorticity-commute}
\end{equation}
Thus, the vorticity is advected by the total velocity $(\mathbf{u}+\mathbf{v})$ and is stretched by the total velocity gradient. In this form one recognizes that the singular vortex filament solutions of (\ref{vorticity-commute}) will move with the total velocity $(\mathbf{u}+\mathbf{v})$, instead of the Biot-Savart velocity $(\mathbf{u}={\rm curl}^{-1}\bom)$ alone.  

\smallskip
\noindent
Note that adding geometric dissipation to the vorticity equation as in equation (\ref{vorticity-commute}) does not require an additional boundary condition for fixed boundaries. 
\end{remark}

\begin{proposition}
The GGF  equation (\ref{vortdiss}) for vorticity including both inertia and dissipation takes the same form as the Euler vorticity equation in two dimensions, but with a modified stream function.
\end{proposition}

\begin{proof}
By a standard calculation with stream functions in two dimensions, equations (\ref{uandv-3D}) and (\ref{Lie-der-GOP-vortex}) imply the following dynamics, expressed in terms of $\omega:=\bhz\cdot\bom$ and $\mu:=\bhz\cdot\boldsymbol\mu$
\begin{equation}
\partial_t\,\omega
+ \big[ \omega,\,
\psi - [\mu[\omega],\tilde{\psi}]\,\big]
=
0
\,,
\label{vortdiss-2D}
\end{equation}
where $\psi=\delta H/\delta\omega$, 
$\tilde{\psi}=\delta E/\delta\omega$ and $[ f,\,g]$ is the symplectic bracket, given for motion in the $(x,y)$ plane by the two-dimensional Jacobian determinant, 
\rem{
\begin{equation}
\big[ f,\,g\big]
=
\frac{\partial f}{\partial x} 
\,\frac{\partial g}{\partial y} 
-
\frac{\partial g}{\partial x} 
\,\frac{\partial f}{\partial y} 
=
\bhz\cdot\nabla f \times \nabla g
\,.
\label{bracket-2D}
\end{equation}
}
\begin{equation}
\big[ f,\,g\big]dx\wedge dy
=
df\wedge dg
\,.
\label{bracket-2D}
\end{equation}
Equation (\ref{vortdiss-2D}) takes the same form as Euler's equation for vorticity, but with a modified stream function, now given by the sum $\psi - [\,\mu,\,\tilde{\psi}\,]$. 
This proves the proposition.
\end{proof}

\begin{remark}
The GGF equation for vorticity in two dimensions (\ref{vortdiss-2D}) recovers equation (4.3) of Vallis et al. \cite{VaCaYo1989} when one chooses $\mu=\alpha\omega$ for a constant $\alpha$ and $E= \frac{1}{2}\int \omega\,\psi \,dxdy$. However, for this choice of mobility, $\mu$,  point vortex solutions are excluded. 
\end{remark}

\begin{proposition}
The GGF  equation for vorticity in two dimensions (\ref{vortdiss-2D}) possesses point vortex solutions, with any choices of $\mu[\omega]$ and $\tilde{\psi}$ for which 
$K=\psi -[\mu\left[\omega\right],\tilde{\psi}]$ is sufficiently smooth.
\end{proposition}

\begin{proof}
Pairing equation (\ref{vortdiss-2D}) with a stream function
$\eta$ yields
\begin{equation}
\langle\, \eta,\,\partial_t{\omega}\,\rangle
=\left\langle\, 
\big[\,\eta,\,K\left[\omega\,\right] \big],\,
\omega\,\right\rangle
\quad\text{ where }\quad
K\left[\omega\right]
=
\psi -[
\mu\left[\omega\right],\tilde{\psi}
]
\,.
\label{omega-psi-eqn}
\end{equation}
Inserting the expression
\[
\omega(x,y,t)=\Gamma(t)\,\delta(x-X(t))\,\delta(y-Y(t))
\]
into the previous equation and integrating against a smooth test function yields
\[
\dot{\Gamma}\,\eta  + \Gamma\, \dot{X}\,\frac{\partial \eta}{\partial X}
 + \Gamma\, \dot{Y}\,\frac{\partial \eta}{\partial Y}
 =
 \Gamma\,
 \frac{\partial \eta}{\partial X}\,
 \frac{\partial {K}}{\partial Y}
 -
 \Gamma\,
 \frac{\partial {K}}{\partial X}\,
 \frac{\partial \eta}{\partial Y}
 \,,
\]
where $\eta$ and $K$ are evaluated at the point $(x,y)=(X(t),Y(t))$.
Thus, the point vortex solutions for equation (\ref{vortdiss-2D}) on the $(X,Y)$ plane satisfy
\begin{equation}
\dot{\Gamma} =0\,,
\qquad\quad
\dot{X} = \frac{\partial {K}}{\partial Y}\,,
\qquad\quad
\dot{Y} = -\,\frac{\partial {K}}{\partial X}
\,,
\label{Ham-2D}
\end{equation}
whose solutions exist provided the function $K$  is sufficiently smooth.
\end{proof}

\begin{remark}
Solutions of the symplectic Hamiltonian system (\ref{Ham-2D}) extend for the case of evolution of arbitrary many point vortices  for the GGF vorticity equation (\ref{vortdiss-2D}) in two dimensions. These solutions represent a set of $N$ vortices at positions $(X_k(t),Y_k(t))$ ($k=1,\ldots,N$) moving in the plane. Properties of the corresponding point vortex solutions of Euler's equations in the plane are discussed for example in \cite{Sa1992}.
\end{remark}

\begin{remark}[Preservation of fluid vorticity properties in 3D]$\quad$\\
As a consequence of the modified vorticity equation (\ref{vorticity-commute}) in commutator form, one easily checks the following properties.
\begin{itemize}
\item
{\bfi Ertel's theorem} is satisfied by the vector field $\,\,\bom\cdot\nabla$ associated to vorticity. By using the commutator notation and the material
derivative $D/Dt$, one can write
\begin{equation}
\frac{D\alpha}{Dt}
:=
\frac{\partial\alpha}{\partial t}+\left({\bf u+v}\right)\cdot\nabla\alpha
=
\bom\cdot\nabla\alpha
\,,\quad\text{ so that } \quad
\left[\frac{D}{Dt}\,,\,\bom\cdot\nabla\,\right] \alpha
=0
\,,
\end{equation}
for any scalar function $\alpha({\bf x},t)$. 
\item
An analogue of the Kelvin's circulation theorem holds for equation (\ref{vorticity-commute}). Upon expressing the vorticity as $\bom={\rm
curl }\,\bf u$, one writes the following dissipative form of the Euler equation for the velocity $\bf u$
\begin{equation}
\partial_t {\bf u+(u+v)\cdot\nabla u}
+u_j\nabla v^j
=-\nabla p
\,,\qquad
\nabla \cdot \mathbf{u} = 0 \, ,
\end{equation}
where $\mathbf{v}$ is given in (\ref{uandv-3D}). This equation may also be expressed as 
\begin{equation}
\partial_t {\bf u+u\cdot\nabla u}
+ \nabla \Big(p 
- \mathbf{u}\cdot\mathbf{v} \Big)
= 
\underbrace{\
{\bf v \times {\rm curl}\, u }\
}_{\hbox{\rm Vortex force}}
\,,\qquad
\nabla \cdot \mathbf{u} = 0 \,,
\label{GOP-vortexforce}
\end{equation}
by using a vector identity.
An equivalent alternative is the Lie derivative form,
\begin{equation}
\left(\partial_t+\pounds_{\bf u+v}\right)\left(\bf u\cdot \textnormal{d}x\right)=-\,\textnormal{d}p
\,.
\end{equation}
Hence, we find that a {\bfi modified circulation theorem} is satisfied,
\begin{equation}\label{euler-dissip}
\frac{d}{dt}\,\,\oint_{\mathcal{C}\left(\bf u+v\right)} \!\!\!\!\!\bf u\cdot \textnormal{d}x\,\,=\,\,0
\end{equation}
for a loop $\mathcal{C}(\bf u+v)$ moving with the ``total'' velocity $\,\bf u+v$. That is, two velocities appear in the modified circulation theorem. One is the ``transport velocity'' $\,\bf u+v$ and the other is the ``transported velocity'' $\,\bf u$.
\item
From equations (\ref{euler-dissip}) and (\ref{Lie-der-GOP-vortex}) one
checks that
\begin{equation}
\left(\partial_t+\pounds_{\bf u+v}\right){\bf\left(\bom\cdot\textnormal{d}S\wedge
u\cdot\textnormal{d}x\right)
=
-\,\bom\cdot\textnormal{d}S\wedge\textnormal{d}}p
=
-\textnormal{div}\left(p\,\bom\right)\textnormal{d}^3\bf x
\end{equation}
so that the {\bfi helicity of the vorticity $\bom$ is conserved}
\begin{equation}
\frac{d}{dt}\,\int\!\!\!\!\int\!\!\!\!\int_\textnormal{Vol}\! \bom\cdot\bf u \,\,\textnormal{d}^3 x\,=\,0
\end{equation}
\end{itemize}
We may summarize these remarks as follows:
\begin{quote}\textsl{%
All of these classical geometric results for ideal incompressible fluid mechanics follow for the modified Euler equation. These results all persist (including preservation of helicity) when transport velocity is replaced as $(\mathbf{u}+\mathbf{v})\to\mathbf{v}$.
}
\end{quote}
\end{remark}

\begin{remark}[Relation of the GGF vorticity equations to Craik-Leibovich theory]$\quad$\\
The Craik-Leibovich (CL) equations \cite{CrLe1976} describe the dynamics of the Eulerian mean fluid velocity $\mathbf{u}$
depending on time $t$ and spatial position $\mathbf{x}$ in three dimensions, when the fluid motion is driven by rapidly oscillating surface waves due to the wind. These circumstances may generate Langmuir circulations - sets of vortices with axes nearly parallel to the wind direction which sometimes occur in the upper layers of
lakes and oceans. \smallskip

In the CL theory, the rapidly oscillating waves at the surface are assumed to be unaffected by the more slowly changing currents below. The effect of the waves on the Eulerian mean flow is parameterized in the CL theory by introducing into the
Navier-Stokes equations a ``vortex force," expressed in terms of a
prescribed Stokes drift velocity, $\mathbf{u}_S(\mathbf{x},t)$. \smallskip

The CL equations are given by,
\begin{equation}
\frac{\partial \mathbf{u}}{\partial t} 
+ \left(\mathbf{u} \cdot \nabla \right) \mathbf{u} 
+ \nabla \varpi =
\underbrace{\
\mathbf{u}_S \times {\rm curl} \, \mathbf{u}\
}_{\hbox{\rm Vortex force}}
+\ 
\nu\Delta \mathbf{u} \, ,
\qquad
\nabla \cdot \mathbf{u} = 0 \, ,
\label{CL1}
\end{equation}
where $\varpi$ is a pressure enforcing incompressibility and $\nu$ is viscosity, ignored hereafter. Ones sees that the GGF vorticity equation (\ref{GOP-vortexforce})  is in the same form as the Craik-Leibovich (CL) equation (\ref{CL1}). However, the velocities in the vortex force in the two cases differ. In both cases, the vortex force may be interpreted as a noninertial force arising from having transformed into a prescribed moving frame. For additional references and discussions of the properties of the CL equations, see also \cite{Ho1996}.

\end{remark}

This completes our investigation of GGF  vortex dynamics. An obvious extension would be to consider GGF vortex patches in two dimensions. For example, we leave to another study the investigation of the {\bfi selective decay hypothesis} of Matthaeus and Montgomery \cite{MaMo1980} for the relaxation to vortex equilibrium states. Instead of pursuing such GGF vorticity considerations further, we shall next turn our attention to the geometric gradient-flow (GGF) equation for an {\it arbitrary} geometric quantity. In particular, we shall perform explicit computations of GGF equations for scalars, one-forms and two-forms. In each case, the GGF evolution equations allow singular (measure-valued) solutions. 
Thus,  the connecting theme of the discussions below with the Euler equations is the presence of singular solutions in GGF equations, akin to point vortices, vortex lines or vortex sheets for the Euler equations, but with geometric order parameters that transform differently from vorticity under smooth invertible maps.

\section{Singular GGF solutions for scalars}
\label{sec:scalars} 
We have seen in section~\ref{sec:GOP-eqn} how GGF equations (\ref{GOP-eqn-brkt})
always allow for singular solutions of the form
\begin{equation} 
\label{clumpons} 
\kappa(\bx,t)=\sum_{a=0}^N \int p_a(s,t)\, \delta\! \left(\bx - \mathbf{q}_a(s,t) \right)\,ds
\,.
\end{equation}
To derive the equations for $p(s,t)$ and $\mathbf{q}(s,t)$, we
substitute the solution ansatz (\ref{clumpons}) into the GGF equation  (\ref{GOP-eqn-brkt}) and integrate it against a smooth test function $\phi$ dual to $\kappa$, thereby obtaining (\ref{calc-sing}). This section illustrates
one particular example.  Two classes of geometric quantities admitting singular solutions of (\ref{GOP-eqn-brkt}) are known \cite{HP-GOP-2007}. One class includes, for example, scalars, 1-forms and 2-forms, and gives characteristic equations for which the characteristic velocity is a nonlocal vector function. A second class, which includes  densities (three-forms on $\mathbb{R}^3$), is a nonlinear nonlocal gradient flow equation (\ref{HP-kappa}).
Since our main interest lies with the first case, we shall
concentrate on the nonlocal characteristic equations.
These characteristic equations have interesting mathematical and
physical properties and many potential applications. 

The fundamental example is an {\it active} scalar, for which $\kappa=f$ is a function. The evolution of a scalar
by equation (\ref{GOP-eqn-brkt}) obeys
\begin{equation}
\partial_t\,f =
 -\,\pounds_{(\mu[f]\diamond\frac{\delta E}{\delta f})^\sharp}f
= -\,\Big(\frac{\delta E}{\delta f}\nabla \mu[f] \Big)^\sharp\cdot\nabla f
\,.\label{scalareq} 
\end{equation}
Equation (\ref{scalareq}) can be rewritten in characteristic form as 
\begin{equation}
df/dt=0
\quad\hbox{along}\quad
d\mathbf{x}/dt
=
\Big(\frac{\delta E}{\delta f}\nabla \mu[f] \Big)^\sharp
\,.
\end{equation}
The characteristic speeds of this equation are {\em nonlocal}
when $\delta E/\delta f$ and $\mu$  are chosen to
depend on the {\em average value}, $\bar{f}$. It is interesting that such problems arise commonly in the theory of quasi-geostrophic convection and may lead to the development of  singularities in finite time \cite{Geostrophic, Chae2005, Cordoba2005}. 

\rem{ 
One may verify that the scalar equation (\ref{scalareq}) admits weak solutions
(\ref{clumpons}). Figure~\ref{fig:scalarevolution} 
shows the spatio-temporal
numerical evolution of $H*f$ given by (\ref{scalareq}) with initial conditions of
the type (\ref{clumpons}) with  $\delta$-functions whose strengths are random
numbers between $\pm 1/8$. We have taken $\delta E/\delta f=H*f:=\bar{f}$ where
$H$ is the inverse Helmholtz operator $H(x)=e^{-|x|/\alpha}$ with $\alpha=1$ and mobility $\mu[f]=\exp(1-\bar{f}^2)$. We
see the evolution of sharp ridges in $\bar{f}=H*f$, which corresponds to
$\delta$-functions in the solutions $f(x,t)$. \smallskip
} 

Explicit equations for the evolution of strengths $p_a$ and
coordinates $\mathbf{q}_a$ for a sum of $\delta$-functions in (\ref{clumpons}) may be derived using (\ref{calc-sing}) when $\mu[f]=H*f=\bar{f}$. The singular solution parameters satisfy 
\begin{eqnarray}
\frac{\partial p_a(t,s)}{\partial t} 
&=& 
{p}_a(t,s) 
\,{\rm div}\,
\Big(\frac{\delta E}{\delta f}
\nabla \mu[f]\Big)^\sharp
\bigg|_{\bx = \bq_a(t,{s})}
\label{weak-fsoln-peqn}\\
{p}_a(t,s) \, \frac{\partial \mathbf{q}_a(t,s)}{\partial t} 
&=&
{p}_a(t,s)\, 
\Big(\frac{\delta E}{\delta f}
\nabla \mu[f]\Big)^\sharp
\bigg|_{\bx = \bq_a(t,{s})}
\label{weak-fsoln-qeqn}
\end{eqnarray}
for $a=1,2,\dots,N$. For the choice $\mu[f]=\bar{f}$, a solution containing a {\it single} $\delta$-function satisfies
$\dot{p}=-A p^3$, so an initial condition $p(0)=p_0$, evolves
according to $1/p(t)^2=1/p_0^2+4\alpha^2 t$. The comparison of $1/p^2$ from numerics with this theoretical
prediction is shown in Fig.~\ref{fig:ampevolution}. 
\remfigure{ 
\begin{figure} [h]
\centering 
\includegraphics[width=6in]{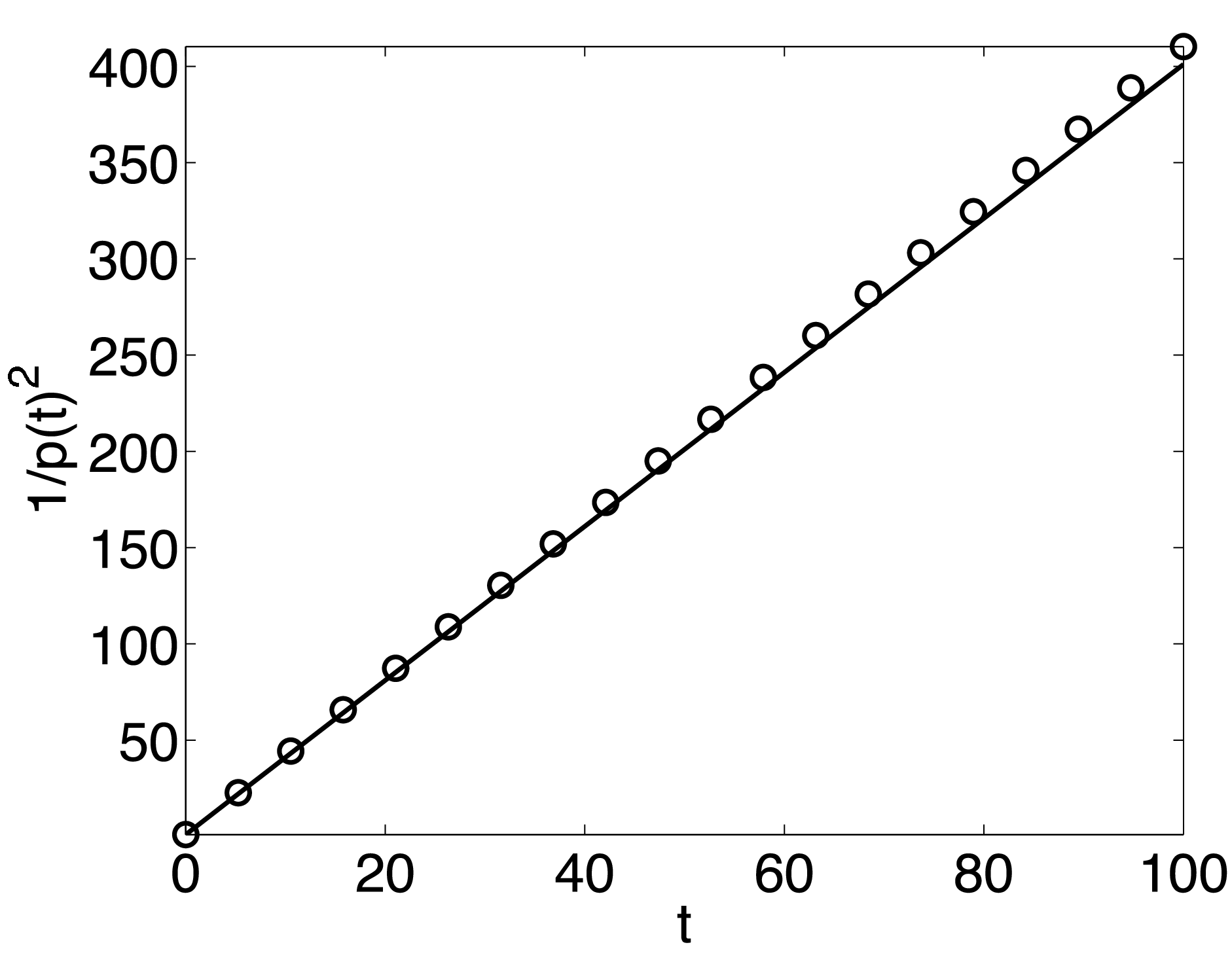} 
 \caption{ 
Evolution of the $\delta$-function strength $1/p(t)^2$ versus time (circles) from \cite{HP-GOP-2007}.
The theoretical prediction $1/p_0^2+4 t$ is shown as a solid line obtained
without any fitting parameters. 
\label{fig:ampevolution} 
 } 
 \end{figure} 
}

\section{Active one-forms and two-forms}
\label{sec:1forms} 
In more generality, one may develop a nonlocal characteristic
equation for the evolution of any geometric quantity. We write the explicit form of the equations for the examples of differential 1-forms
$\mathbf{A} \cdot \mbox{d} \mathbf{x}$ and 2-forms $\mathbf{B} \cdot \mbox{d} \mathbf{S}$  in  three-dimensional space. For this, we begin by computing the the Lie derivative and the diamond operation for these cases. In Euclidean coordinates, the Lie derivatives for these two choices of $\kappa$ are:
\begin{eqnarray}
%
-\pounds_\mathbf{v}\,(\mathbf{A}\cdot d\bx)
&=&
-\left((\mathbf{v}\cdot\nabla)\mathbf{A}+A_j\nabla v^j\right)
\cdot d\bx
\nonumber \\
&=& \left(\mathbf{v}\times{\rm curl}\,\mathbf{A}
-\nabla(\mathbf{v}\cdot\mathbf{A})\right)\cdot d\bx\,,
\nonumber \\
-\pounds_\mathbf{v}\,(\mathbf{B}\cdot {d\bS})
&=&
-\,d \big(v\contract (\mathbf{B}\cdot {d\bS})\big)
-\,v\contract d (\mathbf{B}\cdot {d\bS})
\nonumber \\
&=& 
-\,d \big((\mathbf{v}\times\mathbf{B})\cdot d\bx\big)
-\,v\contract ({\rm div}\,\mathbf{B} d\,^3\,x)
\nonumber \\
&=& \left({\rm curl}\,(\mathbf{v}\times\mathbf{B})
- \mathbf{v}\,{\rm div}\,\mathbf{B}
\right)\cdot {d\bS}
\nonumber
\label{w-eqn}
\end{eqnarray}
Both of these expressions are familiar from fluid dynamics, particularly magnetohydrodynamics (MHD).

From these formulas for Lie derivative in vector form and the definition of diamond in equation (\ref{diamond-def}), we compute explicit expressions for the diamond operation with 1-forms and 2-forms,
\begin{eqnarray}
\left\langle
\boldsymbol\mu[\mathbf{A}]\diamond \frac{\delta E}{\delta \mathbf{A}}
\,,\,\mathbf{u}\right\rangle 
&=&
\left\langle
\frac{\delta E}{\delta \mathbf{A}} \times\,{\rm curl}\,\boldsymbol\mu[\mathbf{A}]
-\, \boldsymbol\mu[\mathbf{A}]\,{\rm div}\,\frac{\delta E}{\delta \mathbf{A}} \,,\,\mathbf{u}\right\rangle
\nonumber \\
\left\langle
\boldsymbol\mu[\mathbf{B}]\diamond \frac{\delta E}{\delta \mathbf{B}}
\,,\,\mathbf{u}\right\rangle 
&=&
\left\langle
\boldsymbol\mu[\mathbf{B}] \times
\,{\rm curl}\,\frac{\delta E}{\delta \mathbf{B}} 
- \, \frac{\delta E}{\delta \mathbf{B}} \, {\rm div}\,\boldsymbol\mu[\mathbf{B}]
\,,\,\mathbf{u}\right\rangle
\nonumber
\end{eqnarray}
for any vector field $\mathbf{u}$. 
\begin{remark}
These formulas result from computing the diamond operation using the standard action of Lie derivatives on differential forms, instead of the action used earlier to calculate equation (\ref{diamond-vectot-pot}) which employed vector potentials for divergenceless vector fields.
\end{remark}
The explicit forms of these equations are unfamiliar and their \rem{vector} solutions will be investigated elsewhere in the context of their
potential applications. Here, we discuss the cases of  exact one- and two-forms.
In these cases ${\rm curl}\,\boldsymbol\mu[\mathbf{A}]=0={\rm div}\,\boldsymbol\mu[\mathbf{B}]$, so the equations simplify considerably. In vector notation one now has
\begin{eqnarray}\label{GOP1}
\frac{\partial \mathbf{A}}{\partial t}
&=&
\text{\large $\nabla$}\left(\left(\boldsymbol\mu[\mathbf{A}]\,\,{\rm div}\,\frac{\delta E}{\delta\mathbf{A}}\right)^{\!\sharp}\cdot\,\mathbf{A}\right)
\\
\label{GOP2}
\frac{\partial \mathbf{B} }{\partial t}
&=& 
{\rm
curl}\left(\left({\boldsymbol{\mu}}[\mathbf{B}]{\times}\,{\rm curl}
\,\frac{\delta E}{\delta \mathbf{B}}\right)^{\!\sharp}
\boldsymbol{\times}\mathbf{B}\right)
\end{eqnarray}

Now, upon defining the vector field
\[
{\bf v(x)}:=\!
\left(
\boldsymbol\mu[\boldsymbol\kappa]\diamond\frac{\delta E}{\delta \boldsymbol\kappa}
\right)^{\!\sharp}
\qquad\text{ with }\quad
\boldsymbol\kappa=\bf A,\, B
\]
the following result holds.
\begin{proposition}
The geometric gradient-flow equations (\ref{GOP1}) and (\ref{GOP2}) for closed one-forms $\bf A$  two-forms $\bf B$ have singular solutions of the form (\ref{clumpons}), where
\begin{align}
\dot{\bf q}_a(t,s)
&=\left.\mathbf{v(\bx)}\right\vert_{{\bf \bx = q}_a}
\nonumber\\
\dot{\bf p}_a(t,s)
&=
{\bf 
\mathbf{p}}_a(t,s)\,\left.\left({\bf\nabla\!\!\cdot\!
v(\bx)}\right)\right\vert_{{\bf \bx = q}_a}
-
\left.{\bf\nabla
v(\bx)}
\right\vert_{{\bf \bx = q}_a}
\!\!
\cdot
\,
\mathbf{p}_a(t,s)
\label{pq-eqns-1form}
\end{align}
for closed one-forms $\bf A$ and
\begin{align*}
\dot{\bf q}_a(t,s)
&=\bf v(\bx)
\vert_{{\bf \bx = q}_a}\\
\dot{\bf p}_a(t,s)
&=
\mathbf{p}_a\text{$\!\!^T$}(t,s)\cdot\nabla
\bf v(\bx)\vert_{{\bf \bx = q}_a}
\end{align*}
for closed two-forms $\bf B$.
\end{proposition}
\begin{proof}
Consider equation (\ref{GOP1}) for {\bf A} written as
\[
\partial_t\,{\bf A}=-\bf\pounds_v A=- \nabla\!\left(v\cdot A\right)
\qquad\text{ with }\quad
\bf v 
=
-\left(\boldsymbol\mu\,\,{\rm div}\,\frac{\delta \text{$E$}}{\delta\mathbf{A}}\right)^{\!\sharp}
\]
Pairing this equation with a smooth vector field $\boldsymbol\phi$, 
substituting the singular solution ansatz (\ref{clumpons}), integrating by parts where necessary and matching all terms in $\boldsymbol\phi$
on the two sides yields equations (\ref{pq-eqns-1form}) for the $\bf q$'s and $\bf p$'s.\smallskip

The result for closed 2-forms is proven by noticing that
\[
{\rm curl}\bf\left( v\times B\right)=B\text{$^T$}\!\cdot\nabla v-v\text{$^T$}\!\cdot\nabla B-\left(\nabla\cdot v\right)
B
\qquad\text{ with }\quad
\bf v=\left(\boldsymbol\mu\times{\rm curl}\,\frac{\delta \text{$E$}}{\delta\mathbf{B}}\right)^{\!\sharp}
\]
then following the same steps as for the case of exact 1-forms.
\end{proof}

When considering the GGF equations (\ref{GOP-eqn-brkt}) for differential forms that are not exact, singular solutions also exist, satisfying more complicated relations, depending
on how the diamond operator is expressed in each particular case. One may see this, by following the same procedure.
\smallskip

For exact one- and two-forms, the vector equations above can be reduced to nonlocal
nonlinear scalar characteristic equations of the form (\ref{scalareq}) for
the potentials:. Note that in $\mathbb{R}^3$ (which is of interest to us here) every closed form is exact since $\mbox{curl } \mathbf{A}=\mathbf{0}$ gives $\mathbf{A}= \nabla \psi$ 
for some scalar $\psi$ and $\mbox{div } \mathbf{B}=0$ necessitates $\mathbf B=\mbox{curl } \mathbf{C}$ for some vector $\mathbf{C}$. The characteristic equations for the potentials are derived in the following 
\begin{proposition} The vector equations (\ref{GOP1}) and (\ref{GOP2}) for exact 1-forms ${\bf A}=\nabla\psi$
and exact 2-forms ${\bf B}={\rm curl}\left(\Psi\,\bf\hat z\right)$ are equivalent to
scalar GGF equations of the type (\ref{scalareq}), in terms of the potentials $\psi$ and $\Psi$. 
Specifically, one finds
\begin{equation}
\frac{\partial \psi }{\partial t}= \Big( \frac{\delta E}{\delta\psi}\,
\nabla \vartheta[\psi] \Big)^\sharp
\boldsymbol{\cdot}\nabla \psi
\,,
\label{APsieq}
\end{equation}
and
\begin{equation} 
\label{psieqB2} 
\frac{\partial \Psi}{\partial t}
=  \Big(\frac{\delta E}{\delta \Psi} \nabla \Phi[\Psi] \Big)^\sharp \cdot \nabla \Psi
\,, 
\end{equation}
where one defines $\boldsymbol\mu[\bf A]:=\nabla \vartheta[\psi]$ and $\boldsymbol\mu[{\bf B}]:={\rm curl}\left(\Phi[\Psi]\,\bf\hat z\right)$.
\end{proposition}
\begin{proof}
Inserting the expression ${\bf A}=\nabla\psi$ in eq. (\ref{GOP1}) yields
\begin{align*}
\frac{\partial \psi }{\partial t}&= 
\Big( \boldsymbol\mu[{\bf A]}\,\,{\rm div}\,\frac{\delta E}{\delta\bf
A}
\Big)^\sharp
\boldsymbol{\cdot}\nabla \psi\\
&=
\Big( \nabla \vartheta[\psi]\,\,\frac{\delta E}{\delta\psi}
\Big)^\sharp
\boldsymbol{\cdot}\nabla \psi
\end{align*}
\rem{
a 1-form allows a simplification in the case
$\mathbf{A}=\nabla \psi$: the equation for potential $\psi$ reads,}
with nonlocal ${\delta E/\delta\psi}$ and $\mu[\psi]$.
\rem{
\begin{equation}
\frac{\partial \psi }{\partial t}= \Big( \frac{\delta E}{\delta\psi}
\nabla \mu[\psi] \Big)^\sharp
\boldsymbol{\cdot}\nabla \psi
\,.
\label{APsieq}
\end{equation} 
This equation for the potential $\psi$ has the same nonlocal
characteristic structure as the scalar equation (\ref{scalareq}).
} 

Similarly, the evolution of 2-form fluxes 
$\mathbf{B} \cdot \mbox{d}\bS= B_x \,\mbox{d}y \wedge \mbox{d}z
+B_y\, \mbox{d}z 
\wedge \mbox{d}x+ B_z\, \mbox{d}x \wedge \mbox{d}y$ also simplifies,
\rem{Again,
the evolution for a general flux $\mathbf{B} \cdot d\bS$ is
quite complicated, but it can be simplified  for the case  ${\rm
div}\,\mathbf{B}=0$ when 
}
when $\mathbf{B}=\nabla\Psi\times\bf\hat{z}$ where $\Psi$ only depends on two spatial coordinates $(x,y)$. Then, 
\[ 
\mbox{curl}\, \frac{\delta E}{\delta \mathbf{B}}
=\mathbf{\hat{z}} \frac{\delta E}{\delta \Psi}. 
\] 
and 
\[ 
\mathbf{\boldsymbol\mu[B]} \times \mbox{curl}\, \frac{\delta E}{\delta \mathbf{B}}
=
(\nabla \Phi \times \mathbf{\hat{z}})
\times \mathbf{\hat{z}} 
\,\frac{\delta E}{\delta \Psi} 
= - \, \frac{\delta E}{\delta \Psi} \,\,\nabla \Phi 
\,.
\] 
Equation (\ref{scalareq}) may be written for the stream function
$\Psi$ (removing the curl from both sides of (\ref{GOP2}))  
\begin{equation} 
\label{psieqB} 
\mathbf{\hat{z}}\frac{\partial \Psi}{\partial t}
=
- \, \Big(\frac{\delta E}{\delta \Psi} \nabla \Phi \Big)^\sharp
\times \mathbf{B}
\rem{ \hskip1cm 
\mbox{with} \hskip1cm {\rm div}\, {\boldsymbol{\mu}}[\mathbf{B}] = 0}
\end{equation}
Then, simplification of two
cross products leads to 
\begin{equation} 
\label{psieqB3} 
\frac{\partial \Psi}{\partial t}
=  \Big(\frac{\delta E}{\delta \Psi} \nabla \Phi \Big)^\sharp \cdot
\nabla \Psi. 
\end{equation}

\rem{
In this case, equation (\ref{scalareq}) may be written for the stream function
$\Psi$ in  $\mathbf{B}=\rem{\mbox{curl}\,\Psi\mathbf{\hat{z}} 
=}\nabla \Psi \times \mathbf{\hat{z}}$ as 
\begin{equation} 
\label{psieqB2} 
\frac{\partial \Psi}{\partial t}
=  \Big(\frac{\delta E}{\delta \Psi} \nabla \Phi \Big)^\sharp \cdot \nabla \Psi
\,, 
\end{equation}
when we choose mobility to be $\mu=\mbox{curl}\,(\mathbf{\hat{z}} \Phi) 
=\nabla \Phi\times\mathbf{\hat{z}} $.
}
Hence, choosing ${\delta E/\delta \Psi}$ and $\Phi$ to depend on the average value $\bar{\Psi}$ again yields a nonlocal characteristic equation.
\end{proof}


\begin{remark}
Equations (\ref{APsieq}) and (\ref{psieqB2}) do allow singular $\delta$-like solutions of the form (\ref{clumpons})
for $\psi$ and $\Psi$. These solutions, however, 
lead to $\delta'$-like singularities in the forms $\mathbf{A}$ and $\mathbf{B}$. One may understand this point by deriving the expressions for  $\psi$ and $\Psi$ corresponding to the  \emph{clumpon} solutions of the form  (\ref{clumpons}) for  $\mathbf{A}$ and $\mathbf{B}$. 

For example, taking the divergence of an exact one-form $\mathbf{A}=\nabla\psi$ yields 
$
\nabla\cdot\mathbf{A}=\Delta\psi.
$
Upon using the Green's function of the Laplace operator $G(\mathbf{\bx},\text{\bfi y})=-\left|\mathbf{\bx}-\text{\bfi y}\,\right|^{-1}$, an expression for $\psi$ emerges in terms of $\mathbf{A}$:
\begin{align*}
\psi(\mathbf{\bx},t)&
=-\!\int  \nabla_{\!\mathbf{\bx}'\,}G(\mathbf{\bx},\mathbf{\bx}')\cdot\!\mathbf{A}(\mathbf{\bx}',t)\,d\mathbf{\bx}'.
\end{align*}
Inserting the singular solution (\ref{clumpons}) for $\bf A$ then yields
\begin{align*}
\psi(\mathbf{\bx},t)&=-\sum_i \int\! ds\,  \mathbf{P}_i(s,t)\cdot\nabla_{\!\mathbf{Q}_i}G(\mathbf{\bx},\mathbf{Q}_i(s,t)).
\end{align*}
However, this singular solution for the potential is \emph{not} in the same form as (\ref{clumpons}), since the singularities for $\psi$ do not manifest themselves as  $\delta$-functions. 
\smallskip

A similar procedure applies to the case of exact two-forms $\mathbf{B}(x,y)=\textrm{curl}\!\left(\Psi(x,y)\,\bf\hat{z}\right)$,
so that $\textrm{curl}\,\mathbf{B}=\Delta\left(\Psi\, \bf\hat{z}\right)$. We have 
\[
\Psi(\bx)\,=\,{\bf\hat{z}}\,\cdot\,
\sum_i\int\!ds\, \mathbf{P}_i(s,t)\times \nabla_{\mathbf{Q}_i}G(\mathbf{\bx},\mathbf{Q}_i(s,t)), 
\]
where $\mathbf{Q}$ is in the plane $(x,y)$.\smallskip

Thus, the equations (\ref{APsieq}) and (\ref{psieqB2}) for $\psi$ and $\Psi$
allow for two species of singular solutions. One of them takes the form
(\ref{clumpons}), while the other corresponds to a $\delta$-like solution
of the same form (\ref{clumpons}) for $\mathbf{A}$ and $\mathbf{B}$.

A deeper explanation of this fact can be given in a general context as follows. Consider the advection
equation for an exact form $\kappa=d\lambda$, with potential $\lambda$
\[
\left( \partial_t  +
\,\pounds_{u}\right)
d\lambda=0.
\]
At this point, one remembers
that the exterior differential commutes with the Lie derivative so that the
equation for the potential $\lambda$ is again an advection equation with
the same characteristic velocity
\[
\left( \partial_t  +
\,\pounds_{u}\right)
\lambda=0
\]
At this point, one obtains singular $\delta$-like solutions of the form
(\ref{clumpons}) for both $\kappa$ and $\lambda$ (provided the characteristic velocity $u$ is sufficiently smooth).

\rem{
Thus, the singular solutions of (\ref{APsieq}) and (\ref{psieqB2}) for $\psi$ and $\Psi$ differ from the singular solutions of the form  (\ref{clumpons}) for  $\mathbf{A}$ and $\mathbf{B}$.
}
\end{remark}

\section{Three more examples}
The developments discussed above produce an interesting opportunity for the addition of dissipation to various other continuum equations. Following the introduction of the dissipative Euler equation above,  one could extend the dissipative diamond flows with any type of evolution operator.  This section sketches how one might develop this idea further, by illustrating its application in three more physically relevant examples.

\paragraph{Dissipative EPDiff}
Consider adding geometric dissipation to the Euler-Poincar\'e equation on the diffeomorphisms (EPDiff)  \cite{HoMa2004} for the evolution of a one-form density $m$ defined by
\begin{equation}
 m=\mathbf{m}\cdot d\bx\otimes d^3x
 \,.
\end{equation}
This addition gives the {\it dissipative EP equation},
\begin{equation}
 \partial_t\, m 
 + {\rm ad}^*_{\delta H/\delta m}\, m 
 = 
-\, \pounds_{\!\big( \mu[m]\,\diamond\, {\delta E/\delta m} \big)^\sharp }\,\, m
\,=\,
-\, {\rm ad}^*_{\!\big( \mu[m]\,\diamond\, {\delta E/\delta m} \big)^\sharp }\,\,m
 \,.
 \label{EPDiff-diss}
\end{equation}
When $H[m]$ is the Hamiltonian for the Lie-Poisson theory corresponding to EPDiff, then the vector field $\delta H/\delta m=u$ is the characteristic velocity for the Euler-Poincar\'e equation. For a one-form density $m$, the diamond operation is given by ${\rm ad}^*$, which is equivalent to Lie derivative. That is,
\begin{equation}
 \mu[m]\,\diamond\, {\delta E/\delta m} 
 =
 {\rm ad}^*_{\delta E/\delta m}\,  \mu[m]
 =
 \pounds_{\delta E/\delta m} \, \mu[m]
 \,.
\end{equation}
The further choice $\mu[m]=\alpha m$ for a positive constant $\alpha$ recovers equation (6.10) of Bloch et al.  \cite{BlKrMaRa1996}. When in addition, $\mu[m]=K*m$ for a smoothing kernel $K$, then equation (\ref{EPDiff-diss}) supports singular solutions of the type discussed in Holm and Marsden \cite{HoMa2004}. 


\paragraph{Peakon dynamics for the GGF modified Camassa-Holm equation} 
In one dimension, the GGF version of the EPDiff equation (\ref{EPDiff-diss}) reduces to,
\begin{equation}
\partial_t m +(u+v)m_x
+ 2m(u+v)_x
 =
0
 \,,
 \label{GOP-CH}
\end{equation}
where $u=\delta H / \delta m$ for a specified Hamiltonian $H[m]$.
The other velocity $v$ is given in one dimension by
\begin{equation}
v = \left(\,{\rm ad}^*_{\delta E / \delta m}\,\mu[m]\,\right)^\sharp
= \frac{\delta E }{ \delta m}\,\partial_x\mu[m]
+ 2 \mu[m]\,\partial_x\,\frac{\delta E }{ \delta m}
\,,
\end{equation}
for arbitrary (smooth) choices of $\mu[m]$ and $E[m]$.
Now consider the singular solution form for $m$ given by a sum of $N$ delta functions,
\begin{equation}\label{m-delta1}
m(x,t) = \sum_{i=1}^N p_i(t)\,\delta(x-q_i(t))
\,,
\end{equation}
and take quadratic functionals $H[m]=1/2\,\langle m,G*m\rangle$ and $E[m]=1/2\,\langle m,W\!*m\rangle$ so that
\[
u(x)=G*m=\sum_{j=1}^N \,p_j\,G(x-q_j)
\]
and
\begin{align*}
v(x)&=W\!*m\,\,\partial_x K*m
+ 2\, K*m\,\,\partial_x W\!*m\\
&=
\sum_{j,k=1}^N p_j\,p_k\,\bigg(K(x-q_j)\,\partial_x \!W(x-q_k)+2\,W(x-q_k)\,\partial_x\!
K(x-q_j)\bigg)\\
&:=
\sum_{j,k=1}^N p_j\,p_k\,\mathcal{R}(x-q_j,x-q_{\,k})
\end{align*}
where we have defined $\mathcal{R}$ for compactness of notation.
Substituting the above expressions into the GGF EPDiff equation (\ref{GOP-CH}) and integrating against a smooth test function yields the following relations for time derivatives of $p_i(t)$ and $q_i(t)$: 
\begin{eqnarray} \label{q-eqn} 
\dot{q}_i & = & 
\left.
\Big(u(x) + v(x) \Big)\right\vert_{x=q_i(t)}\\ \nonumber
& = &
\sum_{j=1}^{N}\,\, p_j\,G(q_i-q_j)\,+\sum_{j,k=1}^{N}\,p_j\,p_k\,\mathcal{R}(q_i-q_j,q_i-q_k)
\,, 
\\
\dot{p}_i & = & 
-\,p_i
\left.
\Big(u'(x) + v'(x) \Big)\right\vert_{x=q_i(t)}\\ \nonumber
& = &
\, -\,p_i\sum_{j=1}^{N}\,\,p_j\,\partial_{q_i}\!G(q_i-q_j)\,-p_i\sum_{j,k=1}^{N}\,p_j\,p_k\,\partial_{q_i}\!\mathcal{R}(q_i-q_j,q_i-q_k)
\,,
\label{p-eqn}
\end{eqnarray} 
\rem{ 
where the Hamiltonian $H_N$ is given by
\begin{equation}\label{Ham-pulson}
H_N =  \frac{1}{2}\sum_{i,j=1}^N p_i\,p_j
\left(G(q_i-q_j) + \sum_{k=1}^N p_k\mathcal{R}(q_i-q_j,q_i-q_k) \right)
\,.
\end{equation}
}      
The choices of $H$, $E$ and $\mu$ as functionals of $m$ determine the ensuing dynamics of the singular solutions. 
In particular,  in the case when the velocity $u$ is given by
\begin{equation}
u[m] = \frac{\delta H }{ \delta m} 
= (1-\alpha^2\partial_x^2)^{-1}*m
=  \frac{1}{2}\int e^{-|x-x'|/\alpha}\, m(x'){\rm d}x'
\,,
\quad\hbox{for}\quad
H = \frac{1}{2}\int m\,u[m] \, {\rm d}x
\,.
\end{equation}
Equation (\ref{GOP-CH}) is a GGF version of the integrable Camassa-Holm equation with peaked soliton solutions \cite{CaHo1993}. Nonlinear interactions of $N$ traveling waves of this system may be investigated by following the approach of Fringer and Holm \cite{FrHo2001}. 
 
\paragraph{Dissipative Vlasov dynamics} 
One may also extend the diamond dissipation framework to systems such as the Vlasov equation in the symplectic framework of coordinates and momenta as independent variables. This extension requires the introduction of the Vlasov Lie-Poisson bracket,
defined for phase space densities $f(q,p)\,\textnormal{d}q\wedge \textnormal{d}p$ on $T^*\mathbb{R}^N$ as
\begin{equation}
\{F,H\}(f)=\iint f\,\left[\frac{\delta F}{\delta f}\,,\,\frac{\delta H}{\delta f}\right]_{qp}dq\wedge dp \, . 
\end{equation}
Here the Lie bracket $[\,\cdot\,,\,\cdot\,]_{qp}$ is now given by the canonical Poisson
bracket on Hamiltonian functions. At this point a new definition of the diamond operator is required. This is found by the well known identification of Hamiltonian
vector fields and their generating functions that gives the relation
\begin{equation}
-\pounds_{X_h}\,f=[f,h]_{qp}
\end{equation}
for any phase space density $f(q,p)$ and any function $h(q,p)$. This relation
 identifies
the symplectic Lie algebra action on the Vlasov distribution and the new kind of \emph{symplectic} diamond (which we denote by $\star$) which may be  computed by applying the general definition as
\begin{equation}
\left\langle g\star f\,,\,h\right\rangle
=
\left\langle g\,,\,-\pounds_{X_h\,} f\right\rangle
=
\Big\langle \big [g\,,\,f\big]_{qp}\,,h \Big\rangle
\end{equation}
for any two functions $g$ and $h$. Extending the previous discussions, one
can write the following form of GGF dissipative Vlasov equation,
\begin{equation}
\frac{\partial f}{\partial t}
+
\left[\,f\,,\,\frac{\delta H}{\delta f}\,\right]_{qp}
\,=\,
\left[\,f\,,\,\left[\,\mu(f)\,,\,\frac{\delta E}{\delta f}\,\right]_{qp}
\,\right]_{qp}
,
\label{Vlasov-diss}
\end{equation}
where, in general, the functionals $H$ and $E$ are independent. 
This equation has the same form as the equations for a dissipative class of Vlasov plasmas in astrophysics, proposed by Kandrup \cite{Ka1991} to model gravitational radiation reaction. Kandrup's formulation for an azimuthally symmetric particle distribution is recovered by choosing a linear phase space mobility $\mu=\alpha f$ with positive constant $\alpha$ and taking $E$ to be $J_z[f]$ the total azimuthal angular momentum for the Vlasov distribution $f$. 
\rem{ 
Equation (\ref{Vlasov-diss}) recovers the dissipative bracket formulations of both Kaufman \cite{Ka1984} and Morrison \cite{Mo1984} when $E=H_f-S$, where $S$ is the Vlasov  entropy functional $S=\int f\log{f}$, the quantity $H_f$ is the single-particle Hamiltonian and $\mu[f]=\alpha f$.  For these choices
} 
More generally, if one chooses $\mu[f]=\alpha f$ and $E$ to be the Vlasov
Hamiltonian $H_f$, the dissipative Vlasov equation (\ref{Vlasov-diss})  assumes the {\bfi double bracket} form,
\begin{equation}
\frac{\partial f}{\partial t}
+
\left[\,f\,,\,\frac{\delta H_f}{\delta f}\,\right]_{qp}
\,=\,\alpha\,
\left[\,f\,,\,\left[\,f\,,\,\frac{\delta H_f}{\delta f}\,\right]_{qp}
\,\right]_{qp}
.
\label{Vlasov-diss-BlKrMaRa}
\end{equation}
This is also the Vlasov-Poisson equation in Bloch et al. \cite{BlKrMaRa1996}. However, in contrast to the choices in \cite{BlKrMaRa1996, Ka1984, Mo1984, Ka1991}, the GGF form of the Vlasov equation (\ref{Vlasov-diss}) allows more general mobilities such as $\mu[f]=K*f$ (which denotes convolution of $f$ with a smoothing kernel $K$). The GGF choice has the advantage of recovering the one-particle solution as its singular solution. \smallskip

\paragraph{Dissipative semidirect product dynamics} 
The equations derived above consolidate the idea that any continuum equation in characteristic form,
\[
\left(\partial_t+\pounds_u\right)\kappa=0
\,,
\]
may be modified to include dissipation via the substitution $u\rightarrow u+v$, in which $v$ is the dissipative velocity term expressed in equation (\ref{GOP-eqn-brkt}).
This idea may also be extended to the semidirect product framework 
presented in \cite{HoMaRa1998}, in order to include compressible fluid flows and plasma fluid models such as MHD. Applications of the semidirect product framework for continuum mechanics are beyond the scope of the present work. However, it would be interesting to pursue these applications in future investigations. An immediate application of the semidirect product framework would be an investigation of ideas of  {\bfi selective decay} in the approach to {\bfi topological equilibria} for example in MHD, as first suggested by Taylor \cite{Ta1974} based on work of Woltjer \cite{Wo1958} and later elaborated by Moffat \cite{Mo1985} and others. A possible counterpoint would be to treat the additional helicity-conserving geometric force as a means of driving a magnetic dynamo, rather than relaxing to an equilibrium.

\section{Conclusions and Comments} 
This paper has provided a general method of constructing dissipative evolutionary equations in the form (\ref{HP-kappa}) for a variety of different types of geometric order parameters $\kappa$. The method produces a plethora of fascinating singular solutions for these evolutionary GGF equations. Each GGF equation is expressed as a Lie derivative (\ref{HP-kappa}). Thus, these are all characteristic equations in a certain geometric sense. However, the characteristic velocities in these equations may be {\bfi nonlocal}. That is, the characteristic velocities may depend on the solution in the entire domain. The equations may possess either or both of the following structures: (i) a conservative Lie-Poisson Hamiltonian structure; (ii) a dissipative Riemannian metric structure. The Hamiltonian evolution of a continuum material by the Poisson structure is similar to Lagrangian fluid dynamics, while the dissipative evolution by the  Riemannian metric structure is a geometric version of gradient flow similar to Darcy's Law. The two types of evolution are combined by simply adding the characteristic velocities in their Lie derivatives. \smallskip

Similar types of equations were discussed by Bloch et al. \cite{BlKrMaRa1996} who studied the effects on the stability of equilibrium solutions of continuum Lie-Poisson Hamiltonian systems of adding a type of geometric dissipation that preserves the coadjoint orbits of the Hamiltonian systems. Such equations have the form
\[
\frac{dF}{dt}= \{F,H\} - \{\{F,H\}\} 
\]
for two bracket operations, one antisymmetric and Poisson ($\{F,H\}$) and the other symmetric and Leibnitz ($\{\{F,H\}\}$).  
In contrast, the GGF theory applies to a more general family of such equations with two independent types of energy $H$ and $E$ and two types of brackets, one antisymmetric and Lie-Poisson, the other derived in (\ref{bracketdef}) from thermodynamic principles; so that systems of the GGF form
\[
\frac{dF}{dt}= \{F,H\} - \{\{F,E\}\} 
\]
need not preserve coadjoint orbits, but can have the additional feature that it admit singular solution behavior. The GGF theory has been shown to apply in a number of continuum flows with geometric order parameters, in examples ranging from gradient flows governed by a nonlocal version of Darcy's Law, to dissipative modifications of Euler fluid flows and plasma dynamics. The various types of singular solutions include point vortices, vortex filaments and sheets, solitons and single particle solutions for Vlasov dynamics.  These singular solutions apparently also apply to a recent generalization of Landau's two fluid model of superfluid $HeII$ \cite{Ho2001}. \smallskip

Each of these equations admits singular solutions which lie on invariant solution manifolds. In some cases, the singular solutions emerge from smooth confined initial conditions. In other cases, such emergent behavior does not occur.   It remains an open question to determine whether the singular solutions of a given geometric type will emerge from smooth initial conditions. After they are created, the singular solutions evolve with their own dynamics. Investigations of the interactions of these singular solutions and the types of motions available to them will be discussed elsewhere. The present paper has derived the dynamical equations for these singular solutions in various cases. However, investigations are left to the future to characterize their formation, stability and interactions, the integrability of their interaction dynamics and their fundamental mathematical nature.

\rem{ 
 Answering the last question seems to require an extension of the idea of momentum mappings to the dissipative regime. In equations (\ref{Ham-2D})  and (\ref{q-eqn}-\ref{p-eqn}) for point vortices and solitons, respectively, the singular solution parameters satisfy canonical Hamiltonian equations. Thus, one may understand these cases as the momentum maps as in \cite{MaWe83} and \cite{HoMa2004}, respectively, for the (left) actions of the diffeomorphisms on embedded subspaces (the support sets of the singular solutions). The other cases in which $p$ and $q$ are of different dimensions must be understood differently. The mappings in these cases remain to be characterized. 
} 

\section*{Acknowledgements} 
We are grateful to A. M. Bloch, A. N. Kaufman, J. E. Marsden, F. Otto, T. S. Ratiu and S. Reich for thoughtful remarks and discussions of double brackets in the geometric formulation of dissipation.
The first two authors were partially supported by NSF grant NSF-DMS-05377891. The work of
DDH was also partially supported  by the US Department of Energy, Office of Science, Applied Mathematical Research, and the Royal Society of London Wolfson Research Merit Award. VP acknowledges the support of A. v. Humboldt foundation, the hospitality of Department for Theoretical Physics, University of Cologne, and the European Science Foundation for partial support through the MISGAM program.

\end{document}